\documentclass[journal]{IEEEtran}

\makeatletter
\@ifundefined{pdfshellescape}{}{}%
\makeatother

\usepackage{xcolor}

\colorlet{shadecolor}{yellow}
\usepackage[pdftex]{graphicx}
\graphicspath{{figs/}}
\DeclareGraphicsExtensions{.pdf,.jpeg,.png}

\usepackage{array}
\usepackage{float}
\usepackage{url}
\usepackage{amsmath,amssymb,amsfonts}
\usepackage{algorithm}
\usepackage{algorithmic}
\usepackage[english]{babel}
\usepackage{amsthm}

\theoremstyle{definition}
\newtheorem{definition}{Definition}

\theoremstyle{remark}
\newtheorem{remark}{Remark}
\newtheorem{proposition}{Proposition}

\theoremstyle{definition}
\newtheorem{assumption}{Assumption}
\begin{document}

    \title{Risk Assessments for Evasive Emergency Maneuvers in Autonomous Vehicles}
  \author{Aliasghar Arab,~\IEEEmembership{Member,~IEEE,}
      Milad Khaleghi,~\IEEEmembership{Member,~IEEE,}\\
      and Koorosh Aslansefat,~\IEEEmembership{\ Member,~IEEE,}
\thanks{(e-mail: aliasghar.arab@nyu.edu, aarab@ncat.edu).}}  
\maketitle

\begin{abstract}


This paper presents a systematic verification and validation (V\&V) framework for the Evasive Minimum Risk Maneuver (EMRM) feature in autonomous vehicles, addressing a critical gap in existing safety assessment methods. We introduce the first formally integrated pipeline that unifies Hazard Analysis and Risk Assessment (HARA), System-Theoretic Process Analysis (STPA), and Finite State Machine (FSM) modeling into a single traceable workflow specifically designed for EMRM V\&V. HARA and STPA are combined through a structured hazard-loss mapping to identify hazards and unsafe control actions; an FSM layer captures hazard-to-loss state transitions that neither method models individually; and the unified framework drives automated scenario generation with measurable parameter-space coverage. Applied to a T-junction EMRM case study, the framework guides 1{,}880 RRT-based simulations spanning ego speed, time-to-collision (TTC), and road friction, uncovering a key physical result: the T-junction geometry gives nearly equal difficulty to stopping and to navigating, so the intermediate mitigation mode occupies only 1.9\% of the feasible parameter space. EMRM steering strategies achieve 81\% collision-avoidance rate and reduce mean residual impact speed from 18.9~km/h to 9.0~km/h compared with emergency braking alone, while the framework attains 100\% hazard, UCA, and parameter-space coverage versus $\leq$1\% for traditional methods. These results demonstrate that the integrated HARA-STPA-FSM framework enables high-resolution, traceable EMRM V\&V that is not achievable with any single method in isolation.
\end{abstract}

\begin{IEEEkeywords}
Autonomous driving, HARA, safety engineering, autonomous vehicles.
\end{IEEEkeywords}

\IEEEpeerreviewmaketitle

\section{Introduction}
\IEEEPARstart{M}{inimum} risk maneuvers (MRM) are safety functions designed to minimize the risk of loss or mitigate potential damage during hazardous scenarios in highly automated or autonomous driving contexts. They can be considered as evolutions of traditional safety features, such as Advanced Emergency Braking (AEB), with an emphasis on leveraging emerging technologies and strategies to reduce accidents and improve overall traffic safety~\cite{nayak2024regulatory}. Similarly to AEB, agile controlled maneuvers inspired by professional car drivers can be used as agile MRM to evade or mitigate hazardous situations~\cite{arab2024safepatent, arab2021phd}. Integration of such features into Advanced Driving Assistant Systems (ADAS) or Autonomous Vehicles (AV) requires comprehensive safety V\&V. However, current safety assessment frameworks face several critical limitations that restrict their effectiveness for advanced autonomous driving features, particularly those designed for edge-case scenarios.

The first limitation concerns the lack of formalized integration between STPA and HARA. Although both methods are valuable individually, they are typically used in isolation without a formally defined combined framework~\cite{grimm2018survey, pang2023survey}. This approach reduces scalability, makes it difficult to expand to broader use cases, and creates gaps in the identification of loss scenarios useful for validation and scenario generation. The absence of a standardized integration approach hinders the development of comprehensive safety assessment tools for complex autonomous systems.

\begin{figure}[ht!]
	\centering
	\includegraphics[width=3.35in]{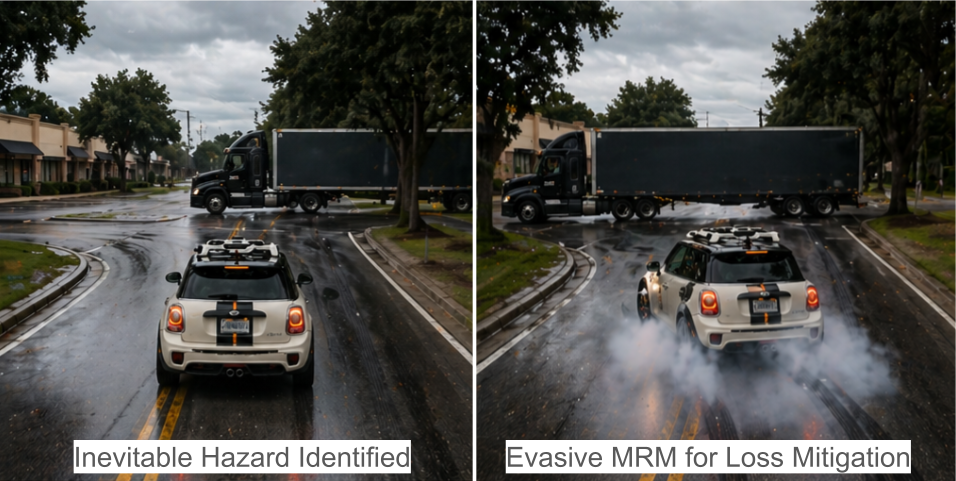}
	\caption{Schematic of an agile autonomous vehicle avoiding a hazardous situation through aggressive maneuvering.}
	\label{fig:realscenario}
 \vspace{-4mm}
\end{figure}

A second limitation involves the inadequate modeling of transitions between hazardous and loss states. Neither HARA nor STPA adequately captures how a system transitions between different hazardous or loss states, which is particularly critical for EMRM features that must respond to evolving unsafe conditions in real-time~\cite{xing2021hazard}. The inability to model state transitions limits traceability and scalability, making it difficult to understand how hazards propagate through the system and how mitigation strategies should adapt.

A third limitation relates to the difficulty in using existing methods for validation and scenario generation. Current methods do not naturally support structured scenario generation or systematic coverage analysis for validation purposes~\cite{arab2023motion, arab2021instructed, han2023safe}. This creates challenges in developing comprehensive test suites and ensuring adequate coverage of hazardous scenarios, particularly for edge-case situations where traditional MRM approaches may be insufficient.

Verification of evasive maneuvers has received considerable attention. Althoff \textit{et al.}~\cite{althoff2010safety} addressed safety verification for coordinated evasive maneuvers; Kianfar \textit{et al.}~\cite{kianfar2013safety} and Pek and Althoff~\cite{pek2019ensuring} studied formal verification and online fail-safe verification for automated driving. Lowe~\cite{lowe2022framework} proposed a framework for emergency obstacle avoidance with validation protocols, while Meltz and Guterman~\cite{meltz2019functional} presented functional safety verification for autonomous systems. Koopman and Wagner~\cite{koopman2016challenges} and Wang \textit{et al.}~\cite{wang2022verification} highlighted challenges and objectives in AV testing and validation. Sun \textit{et al.}~\cite{sun2021scenario} reviewed scenario-based test automation for systematic safety assurance. Despite these advances, a gap remains for high-resolution, scenario-based validation focused on edge-case conditions such as aggressive evasive maneuvers. Our proposed framework addresses this by enabling high-resolution scenario-based validation with explicit focus on edge-case scenarios, including aggressive evasive maneuvers, through integrated hazard-loss modeling and systematic scenario generation.

This paper addresses these limitations by introducing a novel integrated framework that uniquely combines HARA, STPA, and FSM modeling to provide a systematic V\&V methodology specifically designed for the EMRM feature in autonomous vehicles. To the best of our knowledge, this is the first work to formally unify all three methods (HARA, STPA, and FSM) within a single, end-to-end V\&V pipeline for an advanced AV active safety feature. The contribution is threefold: (1)~HARA and STPA are formally integrated through a structured hazard-loss mapping $\mathcal{M}_{HARA\text{-}STPA}$ that builds a traceable link from system hazards to unsafe control actions and loss scenarios; (2)~an FSM-based modeling layer explicitly represents hazard-to-loss state transitions that neither HARA nor STPA models individually, enabling rigorous analysis of how an EMRM feature must respond as the situation escalates; and (3)~the unified framework drives automated scenario generation with measurable parameter-space coverage, replacing ad-hoc test-case selection with a principled, reproducible validation procedure.

The FSM layer is particularly consequential for EMRM V\&V: an evasive maneuver system must not only avoid the hazard when possible, but also gracefully degrade to minimum-damage behavior when avoidance fails. The FSM captures this priority ordering formally through state transitions ($S_3 \xrightarrow{\text{execute}} S_4 \xrightarrow{\text{EMRM\_done}} S_1$ for success; $S_4 \xrightarrow{\text{timeout/error}} S_5$ for controlled-mitigation fallback), enabling the scenario generator to target each transition explicitly rather than discovering failures by chance. This structural coverage (hazard identification, UCA analysis, and state-transition testing in a single traceable workflow) is the key differentiator from prior HARA-only, STPA-only, or simulation-only validation approaches.

To demonstrate the practical application of our integrated framework, consider an emergency vehicle traveling on a highway when suddenly, a truck or deer enters and blocks the road, creating an immediate collision risk, as shown in Fig.~\ref{fig:realscenario}. Emergency vehicles are at high risk for accidents during emergency driving~\cite{weibull2023potential}. Traditional MRM approaches may be insufficient in such edge-case scenarios, where the system must rapidly assess the situation and execute appropriate mitigation strategies. This example illustrates the critical need for a comprehensive safety assessment framework that can handle complex, dynamic scenarios where conventional approaches fall short.

Our proposed framework addresses this challenge by providing a systematic approach to safety assessment that goes beyond traditional methods. Unlike control theory approaches that seek precise performance guarantees through formal analysis, our functional safety assessment framework recognizes the intrinsic limitations of achieving complete formality in complex real-world scenarios and employs a multifaceted strategy to improve the accuracy and comprehensiveness of evaluations~\cite{liu2019safe}. 

The integration of STPA's robustness in identifying systemic hazards through structured analysis~\cite{ishimatsu2010modeling}, FSM's powerful ability to model and analyze state transitions under various conditions~\cite{savelev2021finite}, and HARA's capability to identify hazardous events and describe circumstances leading to specific losses~\cite{suerken2013model} creates a comprehensive safety assessment framework. This unique combination enables systematic identification, modeling, and mitigation of hazards in complex autonomous systems.

The remainder of this paper is organized as follows: Section II provides background on MRM requirements and reviews existing HARA, STPA, and FSM approaches, highlighting their limitations. Section III presents our proposed integrated framework, addressing each of the three challenges systematically. Section IV demonstrates the framework through a case study on EMRM features. Section V presents implementation and analysis validating the framework against the case study. Section VI concludes with a summary of contributions and future directions.

\section{Definitions and Problem Formulation}
\label{sec:ProblemFormulation}
\begin{figure}[t!]
	\centering
	\includegraphics[width=3.3in]{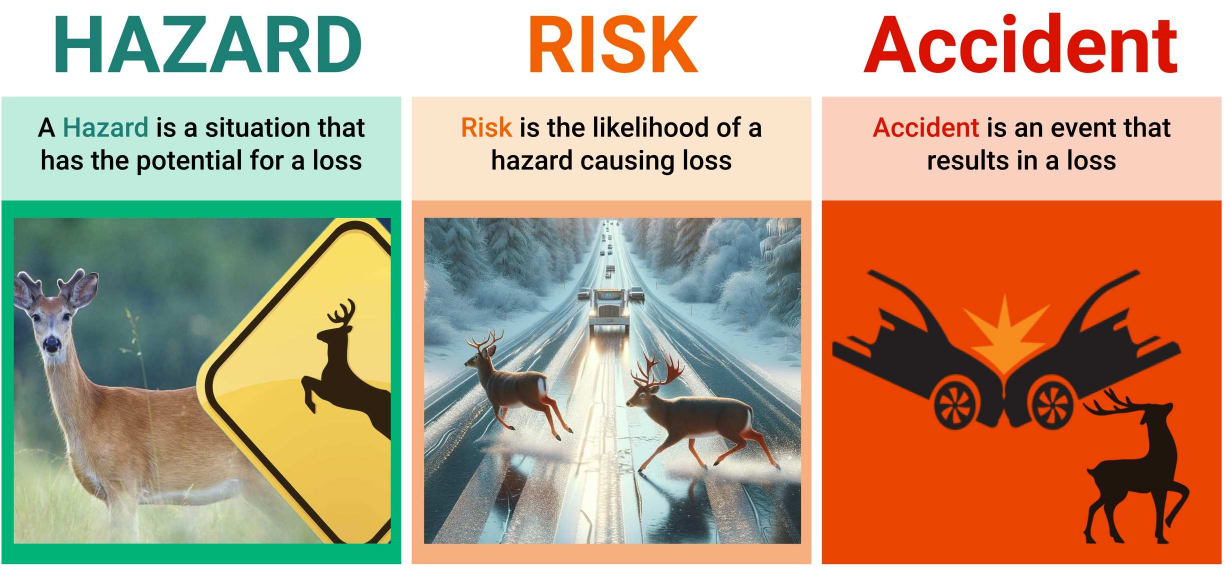}
	\caption{Hazard verses Risks demonstrated for a deer crossing situations.}
	\label{fig:hazardrisk}
 \vspace{-2mm}
\end{figure}

The problem of safety assessment for automated driving features for edge-case scenarios. One of the main challenges in the domain of safety in the automotive industry, especially in the AV industry, is the lack of formal safety assessment methods adapted for advanced functions designed for severe or catastrophic hazards, such as avoiding high-speed obstacles~\cite{ISO26262}. With the growing operational domains of autonomous systems and their interaction with more unpredictable environments, the imperative to ensure the functional safety of MRMs becomes paramount~\cite{UL4600}. Functional safety is fundamentally and simply defined as the minimization of unreasonable risks, while central to this is the development of risk mitigation features for a hazard. Hence, a risk mitigation feature must reduce the probability of an accident or alleviate the harm explained in Fig.~\ref{fig:hazardrisk}. In fact, the MRM feature must guarantee that a risk is mitigated and it's loss is alleviated regardless of where, when, and how the feature is used. In this section, the definitions are based on the above-mentioned standards.

\subsection{World and State Representation}
To establish a formal foundation for safety assessment, we define the world and state representation framework for autonomous vehicles. The world $\mathcal{W}$ represents the complete environment in which an autonomous vehicle operates, encompassing all physical entities, environmental conditions, and dynamic elements that can influence the vehicle's behavior and safety.

\begin{definition}[World Representation]
The world $\mathcal{W}$ is defined as the set of all possible states of the environment, including physical entities (vehicles, pedestrians, infrastructure, obstacles), environmental conditions (weather, lighting, road surface), dynamic elements (traffic patterns, temporal changes), and spatial and temporal relationships between entities.
\end{definition}

\begin{definition}[Vehicle State]
The state $S$ of an autonomous vehicle represents all vehicle and environmental information to represent relative internal and external parameters necessary for autonomous driving.
\begin{equation}
S \in \mathcal{W},
\end{equation}

\end{definition}

\begin{assumption}[State Reconstruction]
Given the vehicle state, $S$, we can reconstruct the vehicle's representation in world $\mathcal{W}$ for verification and validation purposes. This assumption enables state-based analysis for safety assessment, allowing simulation of vehicle behavior, verification of safety properties through state-space exploration, validation of system performance, and generation of test cases based on state transitions.
\end{assumption}

This state-based representation provides the foundation for systematic safety analysis, enabling the identification of hazardous conditions, risk assessment, and the development of appropriate mitigation strategies.

\subsection{Hazard as Risk for Loss}
An autonomous vehicle at any given time $t \in \mathbb{R}>0$ (i.e., positive real time) can be represented by a state $S$, where $S \in W$, and $W$ denotes the set of all possible world states. Although the full set of feasible world states is generally unknown, this formulation separates the world into known and unknown subsets, with the current state $S$ belonging to the known subset.

Any state that has a nonzero risk or probability of transitioning to a loss state is defined as a hazardous state. Let $H \subset \mathcal{W}$ denote the subset of hazardous states, where the vehicle is exposed to a risk of loss. In practice, not all hazardous states can be identified \emph{a priori}, and an unknown subset may exist. Therefore, the set $H$ is inherently dynamic; whenever the risk of harm increases, the corresponding preceding states should be incorporated into $H$.

\begin{definition}[Hazardous State]
A state is considered hazardous if it involves a potential source of harm with a nonzero likelihood of leading to a loss. Such a state is denoted by $S_h$, where
\begin{equation}
s_h \in H.
\end{equation}
The source of a hazard may arise either from internal system malfunctions or from exposure to adverse environmental (world) conditions. This definition is consistent with established terminology in~\cite{ISO26262}.
\end{definition}

Let the subset of $\mathcal{L} \subset \mathcal{W}$ be the set of all loss states where the vehicle is exposed to a potential loss. Classifying these losses according to their severity is necessary to understand if an MRM reduces the risk of severe losses in a high-risk hazardous situation~\cite{arab2024high}. Generalized classification of loss states can be written as 
\begin{align}
\label{eq:hazard1}
\mathcal{L}=\mathcal{L}_1 \cup \mathcal{L}_2 \cup \ldots \cup \mathcal{L}_n,\; i \in [1, n]
\end{align}
\noindent where ranging from the least severe $\mathcal{L}_1$ to the most severe $\mathcal{L}_n$ in with a resolution of $n$ number of loss classes.

\begin{definition}[Loss State]
When the vehicle is in a hazardous state, $S_h$, and an accident occurs, the vehicle's state is transferred to a loss state denoted by $S_l$. The loss state belongs to the set of all loss states.    
\begin{equation}
    S_l\in\mathcal{L}    
\end{equation}
\end{definition}

\begin{definition}[Risk of loss]
The risk of loss is the likelihood that a hazardous state ends up in a loss state in a limited time. A probabilistic function $R_i(S_h) \in [0,\:1]$ can represent the likelihood of that specific incident.
\end{definition}

\begin{definition}[Accident Moment]
The accident moment is defined as the time $t_l$ at which the system state enters the loss set
\begin{align}
\label{eq:hazard2}
S^{t_l}=S_l \in \mathcal{L}.
\end{align}

This marks the final transition from a hazard state to a loss state. 

\end{definition}

\begin{remark}[Reversible vs Irreversible loss]

Distinguishing reversible losses, such as VRU discomfort or minor traffic rule violations, from irreversible losses, such as property damage or human injury, is essential for designing an effective MRM feature. 

We define the overall loss set as $\mathcal{L}$, which is partitioned into two disjoint subsets:
\begin{equation}
\mathcal{L} = \mathcal{L}^{\text{rev}} \cup \mathcal{L}^{\text{irr}}, \quad 
\mathcal{L}^{\text{rev}} \cap \mathcal{L}^{\text{irr}} = \emptyset
\end{equation}
where $\mathcal{L}^{\text{rev}}$ denotes the set of reversible loss states and $\mathcal{L}^{\text{irr}}$ denotes the set of irreversible loss states. A loss state $S_l \in \mathcal{L}^{\text{rev}}$ is reversible when the system can recover to a non-loss state within a finite time horizon without permanent damage or long-term consequences; conversely, $S_l \in \mathcal{L}^{\text{irr}}$ is irreversible when, after entering this state, the system cannot return to any state outside the loss set within a short time horizon, typically due to permanent damage or severe consequences.

To formalize irreversibility using time and state indicators, consider a system state $S_h \in H$ that transitions to a loss state $S_l \in \mathcal{L}_j \subset \mathcal{L}$. If there exists a time $t$ such that
\begin{equation}
{s_h}^t \in \mathcal{H} \;\Rightarrow\; {s_l}^{t+\tau} \in \mathcal{L}^{\text{rev}}, \quad 
\forall t' > t+\tau,\; {s_l}^{t'} \in \mathcal{L}^{\text{irr}},
\end{equation}
then $S_l$ is classified as an irreversible loss state.
\end{remark}

\begin{figure}[ht!]
	\centering
	\includegraphics[width=3.1in]{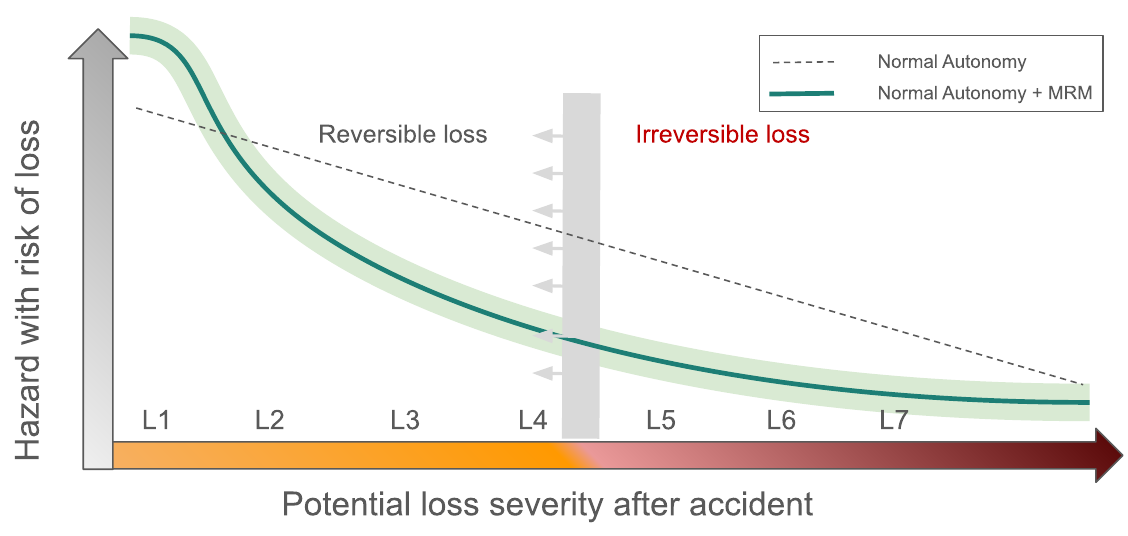}
	\caption{A map of risk of loss at hazardous scenario to potential accident loss with and without MRM.}
	\label{fig:riskloss}
\end{figure}

\subsection{Identifying of hazardous and loss states}
A critical prerequisite for designing an effective MRM functionality is the formal identification of (i) whether the vehicle is currently operating in a hazardous state and (ii) whether an incident has already transitioned the system into a loss state. Recent advances in event-based vision sensors demonstrate that ultra-low-latency hazard perception on public roads is feasible~\cite{gehrig2024lowlatency}. Owing to their microsecond temporal resolution and asynchronous operation, event cameras enable rapid detection of dynamic risk factors, thereby supporting near-immediate classification of the vehicle state as hazardous when applicable.

\begin{assumption}[Hazard Identification] At any given state $S$, the autonomy's perception system can determine the presence of hazards based on the information provided by onboard sensors and perception algorithms.

\end{assumption}
\begin{assumption}[Risk Evaluation:] The risk function $R_i(S)$ can evaluate the probability of hazardous state turning into a loss state ($S_h \rightarrow S_l$) if the existing policy is operating based on predictive algorithms. 
\begin{align}
\label{eq:hazard3}
R_i(S) > \lambda_i \text{ for any } i \in [1, n]
\end{align}
\noindent where $\lambda_i$ represents the risk threshold associated with the transition of the system state from a hazardous state $S_h\in H$ to a loss state $S_l\in \mathcal{L}.$.
\end{assumption}

\noindent Once the system is in state $S$ and the perception module detects a potential hazard, a risk evaluation is performed. If $R_i(S) > \lambda_i$, the system transitions from a hazardous state to a loss state $S_l$. If $S_l \in \mathcal{L}^{\text{irr}}$, an evasive maneuver is required. 


\section{Integrated Safety Assessment Framework}
\label{sec:ProposedMethod}
This section presents our integrated safety assessment framework that addresses the three challenges identified in Section I. The framework systematically combines HARA, STPA, and FSM modeling to provide a comprehensive approach for safety assessment of autonomous driving features, particularly those designed for edge-case scenarios.

\subsection{Framework Overview and Workflow}
The integrated framework provides a systematic approach to safety assessment by Formal Integration of HARA and STPA through standardized procedures and clear definitions with a State Transition Modeling using FSM to explicitly represent hazard and loss state transitions. This unique integrated method allows for the generation of comprehensive test scenarios for coverage analysis. The proposed framework follows a systematic workflow given in Algorithm~\ref{alg:workflow}, where \textit{Step 1} applies the hazard identification function $\mathcal{F}_H$ to context $\mathcal{C}$ to obtain hazardous states $\mathcal{H}$. \textit{Step 2} evaluates potential losses via $\mathcal{F}_L$. \textit{Step 3} identifies UCAs through STPA analysis. \textit{Step 4} applies the integrated mapping $\mathcal{M}_{HARA-STPA}$ to produce hazards, losses, and risk measures. \textit{Step 5} constructs the FSM $G$ capturing state transitions. \textit{Step 6} generates test scenarios using $\mathcal{G}$. \textit{Step 7} executes scenarios and computes coverage metrics.

\begin{algorithm}[ht]
\caption{Integrated Framework Workflow}
\label{alg:workflow}
\begin{algorithmic}[1]
\STATE $\mathcal{H} \leftarrow \mathcal{F}_H(\mathcal{C})$ \quad \textit{// Hazard identification using~\cite{arab2024high}}
\STATE $\mathcal{L} \leftarrow \mathcal{F}_L(\mathcal{C}, \mathcal{H})$ \quad \textit{// Loss assessment}
\STATE $\mathcal{UCA} \leftarrow \text{STPA}(\ldots)$ \quad \textit{// Unsafe control actions (Section III-B.3)}
\STATE $\{\mathcal{H}, \mathcal{L}, \mathcal{R}\} \leftarrow \mathcal{M}_{HARA-STPA}(\mathcal{C}, \mathcal{UCA})$ \quad \textit{// Integration mapping}
\STATE $G \leftarrow (S, \Sigma, \delta, s_0, F)$ \quad \textit{// FSM construction}
\STATE $\mathcal{S}cenarios \leftarrow \mathcal{G}(\mathcal{C}, \mathcal{UCA}, S)$ \quad \textit{// Scenario generation}
\STATE $\text{coverage} \leftarrow \text{Validate}(\mathcal{S}cenarios)$ \quad \textit{// Execute and compute metrics}
\end{algorithmic}
\end{algorithm}

\subsection{Integration of HARA and STPA}\label{sec:HARA-STPA-Integration}
To address the lack of formalized integration between HARA and STPA, we introduce a structured framework that combines these methods with clear definitions and standardized procedures.

\subsubsection{Formal HARA Framework for Edge-Case Scenarios}
We extend traditional HARA with formal definitions specifically designed for edge-case scenarios. An enhanced HARA framework (HARA$^*$) provides structured hazard identification and loss assessment capabilities~\cite{arab2024high}.

\begin{definition}[Formal Hazard Identification]
The HARA$^*$ framework operates in a context $\mathcal{C}$ defined as a structured tuple derived from the world state space $\mathcal{W}$:
\begin{equation}
\label{eq:context}
\mathcal{C} = \left( \mathbf{S}_{t-n:t}, \mathbf{S}_t, \mathbf{S}_{t+1:t+m} \right)
\end{equation}
where $\mathbf{S}_{t-n:t}$ denotes a sequence of prior system/world states (history), $\mathbf{S}_t$ is the current state, and $\mathbf{S}_{t+1:t+m}$ represents a set of feasible future states predicted under the system's dynamics. A \textit{hazard identification function} $\mathcal{F}_H: \mathcal{C} \rightarrow \mathcal{H}$ maps contexts to hazardous states, while a \textit{loss assessment function} $\mathcal{F}_L: \mathcal{C} \times \mathcal{H} \rightarrow \mathcal{L}$ evaluates potential losses associated with each hazard.
\end{definition}

\subsubsection{STPA Implementation}
\begin{figure}[h!]
	\includegraphics[width=3.4in]{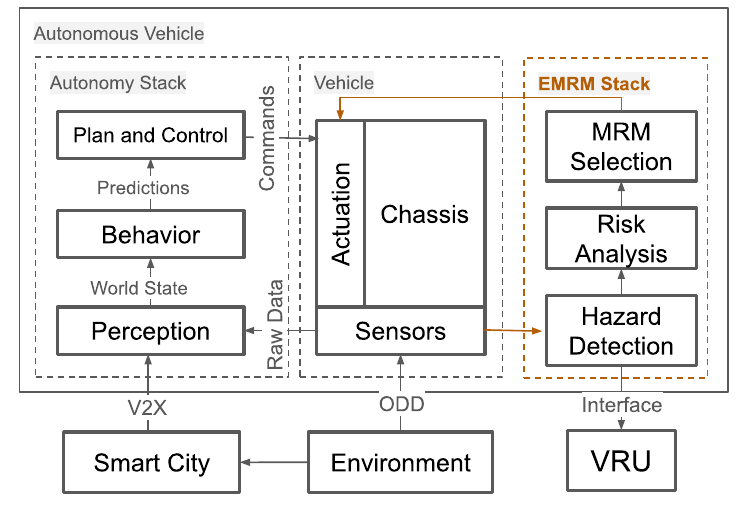}
	\caption{Simple control structure for an autonomous driver with an MRM module.}
	\label{fig:FigSimpleSTPA}
\end{figure}

We integrate STPA to identify unsafe control actions (UCAs) that can lead to hazardous states. This integration creates a formal mapping between system architecture and potential failures. STPA starts by constructing a control structure to analyze potential transitions between hazardous or nonhazardous situations, as shown in Fig.~\ref{fig:FigSimpleSTPA}.

\begin{definition}[Integrated STPA-HARA Framework]
STPA builds on this by identifying how control actions can lead to hazardous states through four specific failure modes~\cite{thomas2013extending}. The integrated framework combines HARA$^*$ hazard identification with STPA control analysis through a formal mapping as
\begin{equation}
\label{eq:integrated_mapping}
\mathcal{M}_{HARA-STPA}: \mathcal{C} \times \mathcal{UCA} \rightarrow \{ \mathcal{H}, \mathcal{L}, \mathcal{R} \}
\end{equation}
where $\mathcal{UCA}$ represents the set of unsafe control actions identified by STPA analysis, and the mapping $\mathcal{M}_{HARA-STPA}$ produces hazards $\mathcal{H}$, losses $\mathcal{L}$, and risk measures $\mathcal{R}$.
\end{definition}



\subsection{FSM for Hazard-Loss Transitions}
\label{sec:FSM-Modeling}
To address the lack of modeling for transitions between hazardous and loss states, we introduce a comprehensive approach based on FSM that explicitly represents how hazards evolve to possible losses.

\begin{definition}[Hazard-Loss FSM]
A Hazard-Loss Finite State Machine is defined as a tuple
\begin{equation}
\label{eq:formalFSM}
G = (\mathbf{S}, \Sigma, \delta, \mathbf{S}_0, \mathbf{S}_F)
\end{equation}
$\Sigma$ is a finite set of events that include hazard detection, risk escalation, mitigation actions, and loss occurrence, $\delta: \mathbf{S} \times \Sigma \rightarrow \mathbf{S}$ is the transition function that defines state changes, $\mathbf{S}_0 \in \mathbf{H}$ is the initial state after the hazard appears (typically detected through hazard perception), and $\mathbf{S}_F \notin \mathbf{H}$ is the final states where the state returns to normal and the hazard no longer exists or an accident has occurred (loss states or back to normal operation).
\end{definition}

\subsubsection{State Transition Modeling}
The FSM explicitly models transitions between different states, enabling traceability and systematic analysis of hazard propagation.

\begin{definition}[State Transition Categories]
We categorize state transitions into three main types:
\begin{enumerate}
\item Hazard Detection Transitions: $s_{normal} \xrightarrow{hazard\_detected} s_{hazardous}$
\item Risk Escalation Transitions: $s_{hazardous} \xrightarrow{risk\_escalation} s_{critical}$
\item Loss Occurrence Transitions: $s_{hazardous} \xrightarrow{mitigation\_failure} s_{loss}$
\end{enumerate}
\end{definition}

\begin{figure}[h!]
	\centering
	\includegraphics[width=3.1in]{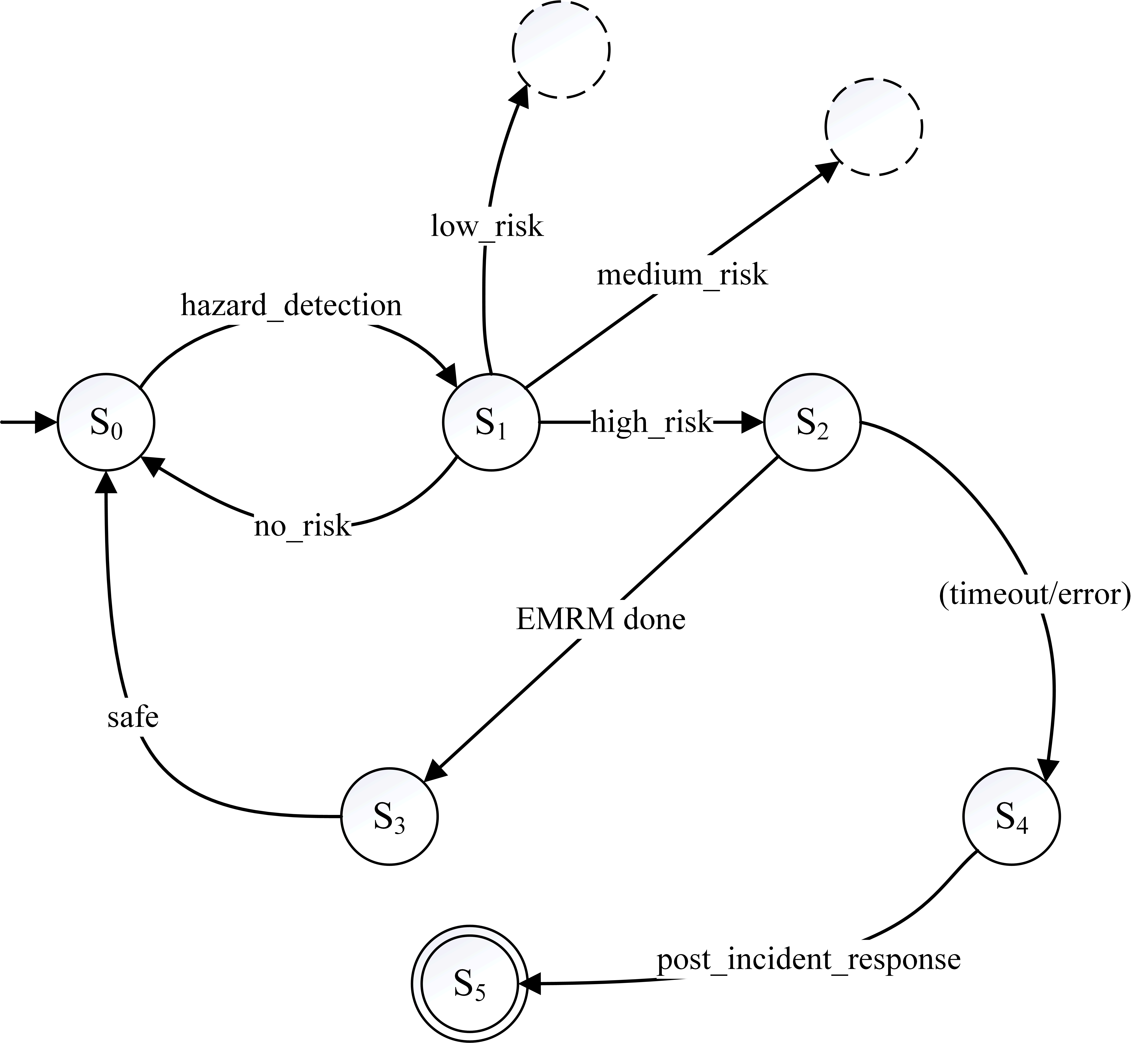}
	\caption{The FSM of EMRM system for an autonomous driving system.}
	\label{fig:FSA for EMRM system}
\end{figure}

\subsubsection{State Transition Analysis}
The FSM explicitly models transitions between states, enabling systematic analysis of hazard propagation. The key transitions include:
\begin{itemize}
\item Detection Transition: $S_1 \xrightarrow{hazard\_detected} S_2$
\item Risk Escalation: $S_2 \xrightarrow{high\_risk} S_3$
\item Maneuver Execution: $S_3 \xrightarrow{execute\_maneuver} S_4$ (via EMRM execution)
\item Success Recovery: $S_4 \xrightarrow{EMRM\_done} S_1$ (return to normal)
\item Failure to Loss: $S_4 \xrightarrow{timeout/error} S_5$ (mitigation failure)
\item Post-Incident: $S_5 \xrightarrow{post\_incident\_response} S_6$
\end{itemize}

The three components (HARA-STPA integration, FSM modeling) work together to provide a comprehensive safety assessment framework to initiate the scenario generation for validations. The scenario generation function $\mathcal{G}: \mathcal{C} \times \mathcal{UCA} \times S \rightarrow \mathcal{S}cenarios$ enables systematic generation of test scenarios covering hazards, UCAs, and state transitions. Coverage metrics include hazard coverage, transition coverage, and UCA coverage. The framework is demonstrated in the following section through a case study with validation.

\section{Case Study}
\label{sec:CaseStudy}
This section demonstrates our integrated safety assessment framework through a case study in which a feature of minimum risk emergency manoeuvring is provided. We apply the framework to two representative scenarios and validate using analytical model-based simulations. An analytical model-based simulation with probabilistic modeling of accidents is implemented. We assume that the impact of an incident can be anticipated using energy loss methods based on relative speeds between the ego vehicle and obstacles or VRUs. Several methods have focused on mitigation of damage alongside avoidance: minimal injury risk motion planning with active mitigation~\cite{guardini2022minimal}, crash mitigation in motion planning~\cite{wang2019crash}, data-driven minimization of collision severity~\cite{parseh2021data}, motion planning with post-impact motions to minimize collision risk~\cite{parseh2023motion}, emergency operation strategies for emergency scenarios~\cite{gong2025emergency}, incorporation of accident risk evolution into motion control~\cite{guo2026incorporating}, and collision dynamics modeling with self-learning control~\cite{chen2025collision}. These approaches support the use of energy-based impact assessment for validation.

The EMRM system is designed to perform aggressive manoeuvres in emergency situations where traditional MRM approaches are insufficient~\cite{arab2024safepatent}. The system must rapidly assess hazardous situations and execute appropriate mitigation strategies to avoid or minimize losses. We apply the proposed framework to two scenarios in which aggressive maneuvering can minimize the risk of inevitable loss. HARA-STPA will be applied to extract losses and hazardous scenarios, and FSM will be implemented to represent state transitions. This framework allows system engineers to validate that an EMRM feature can minimize overall loss. The following examples show how this method can be applied while irreversible losses are prioritized.

\subsection{Critical Scenario at a T-Junction}
\begin{figure}[ht]
	\centering
	\includegraphics[width=3.3in]{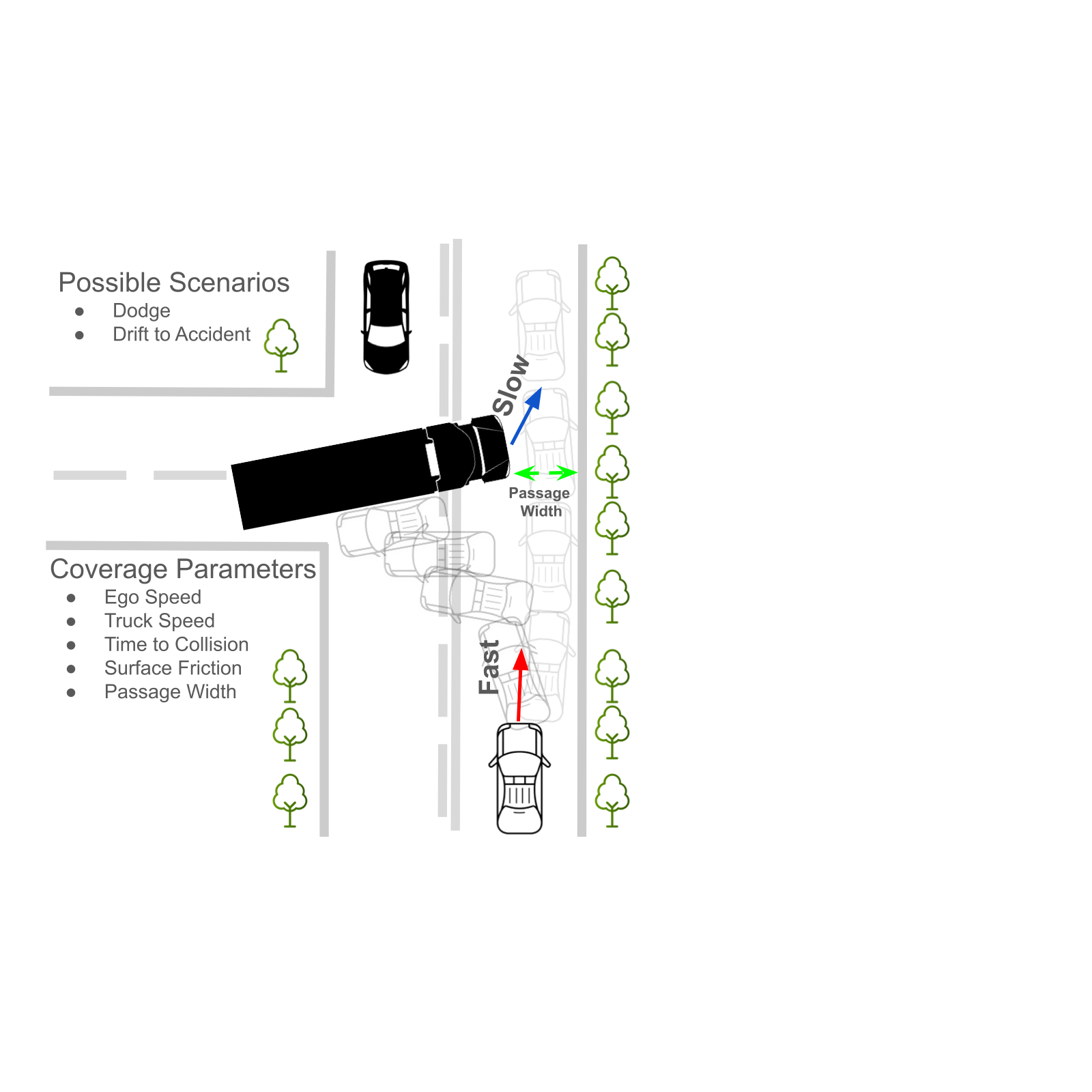}
	\caption{Scenario A: T-junction suburban environment with a large truck blocking the main road.}
	\label{fig:EMRMScene1}
\end{figure}

A T-shaped intersection in a suburban environment. A large truck enters and blocks the main road, making a collision potentially inevitable for an approaching ego vehicle. The hazard arises from the high speed of the ego vehicle and the limited passage width due to the occlusion of the truck on the roadway. Fig.~\ref{fig:EMRMScene1} illustrates the scenario.

\begin{table}[h!]
\centering
\setlength{\tabcolsep}{3pt} 
\begin{tabular}{|>{\raggedright\arraybackslash}p{0.1\columnwidth}|>{\raggedright\arraybackslash}p{0.3\columnwidth}|>{\raggedright\arraybackslash}p{0.45\columnwidth}|}
\hline
SC & Control Action & Hazardous Control Actions (T-Junction) \\ \hline
EMRM & Not Providing Causes Hazard & EMRM fails to execute evasive maneuver when truck blocks roadway; ego vehicle collides with truck or other road users. \\ \cline{2-3} 
& Providing Causes Hazard & EMRM executes aggressive maneuver when no hazard exists; unnecessary evasion may cause rear-end collision or loss of control. \\ \cline{2-3} 
& Wrong Timing or Order Causes Hazard & EMRM initiates maneuver too late (insufficient TTC) or too early; may cause collision or secondary hazards (e.g., oncoming traffic). \\ \cline{2-3} 
& Stopped Too Soon or Applied Too Long & EMRM terminates maneuver prematurely before clearing obstacle, or prolongs maneuver into unsafe lane/position. \\ \hline
Ego AV & \multicolumn{2}{>{\raggedright\arraybackslash}p{0.8\columnwidth}|}{Ego AV is equipped with EMRM and operates in the T-junction scenario with truck blocking the main road.} \\ \hline
\end{tabular}
\caption{Hazardous control actions for T-junction EMRM scenario.}
\label{tab:hazardous_actions}
\end{table}

\subsubsection{Step 1 - Hazard Identification (HARA* Application)}
Following Step 1 of the workflow, we apply the HARA* framework~\cite{arab2024high} to the T-junction scenario. The context $\mathcal{C}_A$ for Scenario A yields hazards such as: truck blocking roadway, limited escape path, and inevitable collision when avoidance is infeasible. As noted in~\cite{arab2024high}, this is definitively a \textit{critical} situation requiring higher-resolution assessments. Accordingly, we split losses into multiple levels (Table~\ref{tab:loss}) and distinguish irreversible losses from reversible ones, as suggested in the problem formulation (Remark on reversible vs.\ irreversible loss in Section~\ref{sec:ProblemFormulation}). The hazard identification function $\mathcal{F}_H$ maps these contexts to hazardous states $\mathcal{H}$. The complete set of scenarios, including the T-junction case, is summarized in Table~\ref{table:list of scenarios}.

\begin{table*}[ht!]
\centering
\caption{List of hazardous scenarios including T-junction and loss reduction evaluations.}
\label{table:list of scenarios}
\begin{tabular}{|c|c|>{\centering\arraybackslash}p{0.61\linewidth}|c|c|c|c|}
\hline
$\mathcal{L}$ID & HID & Evasive MRM Scenario & Manu. & Avoi. & Mit. & SMIL \\ \hline
$\mathcal{L}$8 & H8 & EMRM system performs aggressive maneuvers (dodge or drift) to avoid or mitigate collision when a truck blocks the main road at a T-junction, reducing loss severity for ego and other vehicles. & M$_3$ & A$_3$ & $\mathcal{M}_3$ & D \\ \hline
$\mathcal{L}$1 & H1 &  EMRM system performs aggressive maneuvers to reduce accident impact and loss severity on the VRU due to inability to detect the VRU properly. & M$_3$ & A$_2$ & $\mathcal{M}_1$ & D \\ \hline
$\mathcal{L}$2 & H2 & EMRM system performs hard braking and turning direction to avoid an object due to False-Positive detection. & M$_3$ & A$_2$ & $\mathcal{M}_2$ & B \\ \hline
$\mathcal{L}$3 & H4 & EMRM system performs aggressive maneuvers immediately to reduce accident impact and loss severity for the ego and other vehicles. & M$_3$ & A$_3$ & $\mathcal{M}_3$ & D \\ \hline
$\mathcal{L}$4 & H5 & EMRM perform non-aggressive maneuvers to reduce accident impact due to wrong loss predictions. & M$_3$ & A$_3$ & $\mathcal{M}_2$ & F\\ \hline
$\mathcal{L}$5 & H6 & EMRM system performs hard braking and/or turning direction to avoid an object due to false severity identification. & M$_2$ & A$_1$ & $\mathcal{M}_1$ & F \\ \hline
$\mathcal{L}$6 & H7 & EMRM perform less evasive maneuvers to reduce accident impact due to system limits. & M$_3$ & A$_2$ & $\mathcal{M}_2$ & D\\ \hline
$\mathcal{L}$7 & H9 & EMRM perform maneuvers with delay to reduce accident impact due to system limits. & M$_3$ & A$_2$ & $\mathcal{M}_1$ & D\\ \hline
\end{tabular}
\end{table*}

\subsubsection{Step 2 - Loss Assessment}
Following Step 2 of the workflow, we evaluate potential losses associated with each identified hazard using the loss assessment function $\mathcal{F}_L$. For the T-junction scenario, the relevant loss levels are given in Table~\ref{tab:loss}. The table is structured to separate \textit{reversible} losses (L1--L5) from \textit{irreversible} losses (L6--L7) by a double line, consistent with the problem formulation in Section~\ref{sec:ProblemFormulation}. In the T-junction case, the primary risks are L4 (vehicle damage), L6 (injury to people), and L7 (loss of life) when collision with the truck or other road users occurs.

\subsubsection{Step 3 - STPA Analysis Application}
Following Step 3 of the workflow (STPA Analysis Procedure in Section III-B.3), STPA analysis reveals unsafe control actions that could lead to hazardous states. Table~\ref{tab:hazardous_actions} summarizes the hazardous control actions for the T-junction scenario:
\begin{itemize}
\item UCA1: Not providing an evasive maneuver when hazard is detected
\item UCA2: Providing an evasive maneuver when there is no hazard
\item UCA3: Wrong timing or sequencing of maneuver execution
\item UCA4: Premature termination or excessive duration of maneuvers
\end{itemize}

\subsubsection{Step 4 - Integration Mapping}
\begin{figure}[h!]
	\centering
	\includegraphics[width=3.3in]{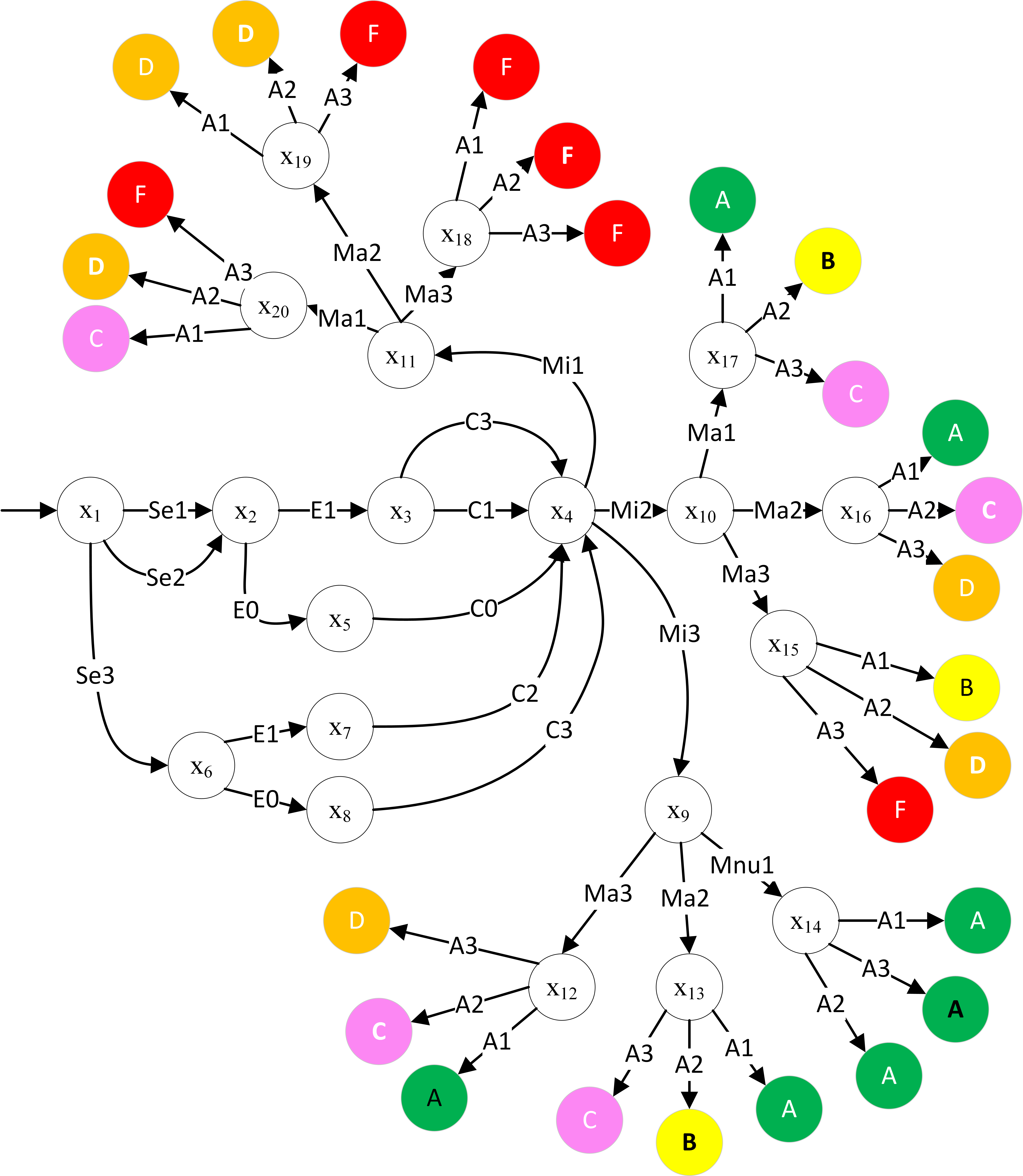}
	\caption{Loss evaluation FSM operating within state $S_3$ of the EMRM FSM.}
	\label{fig:loss evaluation}
\end{figure}

Following Steps 4 and 5 of the workflow, we apply $\mathcal{M}_{HARA-STPA}$ to create formal mappings between the identified hazards, UCAs, and system architecture, and then construct the FSM for the EMRM system following the FSM modeling framework presented in Section III-C. The loss evaluation logic within state $S_3$ is captured by the FSM in Fig.~\ref{fig:loss evaluation}, which determines the appropriate mitigation strategy based on hazard severity, exposure, controllability, and maneuverability. FSM can visualize the state, transitions, and conditions of the system under which these hazardous situations occur and provides a clear map of possible control actions to mitigate losses. This method is an appropriate technique for managing complex systems like AVs equipped with agile maneuvering features to avoid hazards and mitigate losses.

The proposed FSM for the EMRM system, together with its state space and event set, is shown in Fig.~\ref{fig:FSA for EMRM system}. The state set $S$ comprises six states, $S = \{S_i|i=1,\cdots,6\}$. $S_1$ represents the normal driving condition where no obstacles are present. Upon hazard detection, the system transitions to $S_2$, which denotes the identification of an obstacle and the initiation of a risk evaluation process. If the risk is categorized as high, the system enters $S_3$, representing a state where a collision is imminent and an aggressive maneuver must be executed. Risk levels classified as medium or low trigger transitions to other predefined behavioral scenarios not discussed in this paper. Within $S_3$, the system conducts a comprehensive hazard analysis, risk assessment, and loss evaluation to determine the potential reduction in loss severity achievable through the EMRM. If the EMRM is validated and executed successfully, the system transitions to $S_4$, which reflects recovery and a return to safe driving. If the maneuver fails due to a timeout or execution error, the FSM transitions to $S_5$, representing a mitigation failure. This is followed by a transition to $S_6$, where a post-incident response (e.g., alerting emergency services, activating hazard lights) is initiated to reduce residual risk and ensure safety. The state $S_6$ is a terminated state belonging the mark state set, $S_6 \in S_M$.

The input event set $\Sigma$ includes the following events:
\begin{itemize}
    \item hazard-detected: signals that an obstacle has been identified in the vehicle path.
    \item no\_risk: implies that the detected obstacle does not require an evasive action.
    \item high\_risk: indicates that the obstacle presents a significant danger.
    \item EMRM done: confirms that the vehicle has completed an evasive maneuver.
    \item timeout/error: indicates that the evasive maneuver was unsuccessful.
    \item post incident response: implied that the vehicle has completed an aggressive maneuver such as emergency braking or an immediate pull-over.
    \item safe: asserts that the system is now in a safe condition and can resume normal operation.
\end{itemize}

Fig.~\ref{fig:loss evaluation} illustrates the FSM that operates when the EMRM FSM is in state $S_3$, with the purpose of ensuring safety and minimizing losses after the execution of the EMRM system. This FSM has been implemented for the list of scenarios provided in Table~\ref{table:list of scenarios}. The set of events $SE = \{Se_1,Se_2,Se_3\}$ is the main type of hazards, where $Se_1$ is \textit{Marginal} hazard, $Se_2$ is $Critical$ hazard, and $Se_3$ represents $Catastrofic$ hazard. The set of $E = \{E_0,E_1\}$ is the set of $Exposure$ classes, where the event $E_0 $ represents the situation with $Low$ probability of the occurrence of a hazard, and $E_1$ is the situation where the probability of the occurrence of a hazard is $Very\_Low$. The set of $C=\{C_0,C_1,C_2,C_3\}$ shows the different levels of controllability for the EMRM system, where $C_0$ is the $High$ level of controllability, $C_1$ is the $medium$ level, $C_2$ is the $Low$ level, and $C_3$ is the $Very\_Low$ level. The set $MI = \{Mi1,Mi2,Mi3\}$ repesents the level of manuverability of the vehicle, where $Mi1,Mi2$, and $Mi3$ are $Low$, $Medium$, and $High$ level of manuverability, respectively.

\subsubsection{Step 5 - FSM Development}
To develop an FSM for an EMRM system for an autonomous driving system, a tuple $G=(S,\Sigma,\delta,s_{0})$ is considered, where $S$ is a finite set of states; $\Sigma$ is a finite set of events, $\delta:S\times\Sigma\to S$ is the transition function, and $s_{0}$ is the initial state of the system. $S_M \subseteq S$ denotes the set of state markers representing a terminating condition. The concatenation of events forms a string. $\Sigma^{*}$ is the set of all finite strings of events in $\Sigma$, as well as the zero-length string $\epsilon$. We can extend the definition of $\delta$ to strings in $\Sigma^{*}$, by recursively defining $\delta(s,\sigma.e)=\delta(\delta(s,\sigma),e)$, for any string $\sigma \in \Sigma^*$ and any event $e\in\Sigma$, where $\delta(s,\epsilon)=s$. The set $L(G(s)=\{\sigma\in\Sigma^{*}~|~\delta(s,\sigma) \text{ is defined}\}$ contains the strings that can be generated from $s$. For the T-junction scenario, the FSM loss assessment in Fig.~\ref{fig:loss evaluation} operates within state $S_3$ to select the appropriate evasive strategy (avoidance vs.\ mitigation) based on the severity and controllability of the threat.

\subsubsection{Step 6 - Scenario Generation}
Following Steps 6 and 7 of the workflow, we apply the scenario generation function $\mathcal{G}$ to produce test scenarios for the T-junction case. The possible mitigation scenarios for this hazardous situation are summarized in Table~\ref{tab:mitigation_scenarios} (Section~\ref{sec:Results}). Each scenario is characterized by the maneuver difficulty, the probability of loss with no mitigation, and the mitigability success, i.e., the likelihood that the maneuver reduces loss severity. The difference between the original (unmitigated) loss level and the mitigated loss level indicates the benefit of each strategy. \textit{Dodge} (steering to avoid without drifting) is the least demanding maneuver but requires sufficient passage width and TTC. \textit{Drift to avoid} uses controlled drifting to create lateral displacement when grip is limited; it is more difficult but can succeed in tighter spaces. \textit{Drift to accident} represents a last-resort mitigation where collision is inevitable but impact severity is reduced (e.g., glancing vs.\ head-on).

\subsubsection{Step 7 - Coverage Test and Validation}
To determine the parameter ranges over which EMRM can effectively mitigate the T-junction hazard, coverage tests are run in simulation. The key coverage parameters and their influence on mitigability are summarized as follows. Ego speed: higher ego speeds reduce the time available for reaction and increase impact energy; mitigation is feasible within a range where TTC remains above a maneuver-dependent threshold (e.g., $\sim$0.5--1.5\,s for dodge, $\sim$0.3--1.0\,s for drift). Truck speed: the relative speed between ego and truck affects closure rate and TTC; slower truck speeds generally improve mitigability. TTC is critical for maneuver feasibility; below a minimum TTC, only drift-to-accident mitigation may be viable. Surface friction: lower friction reduces lateral grip and increases the difficulty of dodge and drift-to-avoid maneuvers; drift strategies become more relevant on low-friction surfaces. Passage width: the lateral space between the truck and road boundary determines whether dodge (requires wider passage) or drift (can operate in narrower gaps) is feasible. Coverage simulations sweep these parameters to identify the boundaries of the mitigable region; experimental results and analysis are presented in Section~\ref{sec:Results}.


\section{IMPLEMENTATION AND ANALYSIS}
\label{sec:Results}
This section presents the implementation and analysis that validate our integrated framework against the T-junction case study (Section~\ref{sec:CaseStudy}). We implement analytical model-based simulations to duplicate the scenarios derived from the framework (Steps 1--7), enabling quantitative proof of the method. Scenarios are instantiated using C++-based simulation with energy loss models for impact assessment~\cite{parseh2021data, parseh2023motion}.

\subsection{Implementation}
\label{subsec:Implementation}

\subsubsection{Simulation Environment and Analytical Models}
The T-junction scenario (EMRM Scene~1, Fig.~\ref{fig:EMRMScene1}) is replicated in a simulation environment using analytical vehicle dynamics and collision models. The implementation comprises: (i) a scenario-validation framework with EMRM Scene~1 YAML definition for the T-junction truck-blocking case; (ii) an analytical validation script that uses energy loss methods~\cite{guardini2022minimal, wang2019crash} to estimate impact severity from relative speeds between ego vehicle and obstacles. The framework assumes perception (camera, lidar) locates the vehicle and detects objects, road lines, and obstacles. Coverage parameters (ego speed, truck speed, TTC, surface friction, passage width) are swept as in Step 7 of the case study. Table~\ref{tab:mitigation_scenarios} summarizes the mitigation scenarios validated.

\begin{table}[h!]
\centering
\setlength{\tabcolsep}{2pt}
\begin{tabular}{|>{\raggedright\arraybackslash}p{0.12\linewidth}|c|>{\centering\arraybackslash}p{0.24\linewidth}|>{\centering\arraybackslash}p{0.16\linewidth}|>{\centering\arraybackslash}p{0.12\linewidth}|>{\centering\arraybackslash}p{0.16\linewidth}|}
\hline
Mit. Scen. & Diff. & Range & $P$(loss) w/o Mit. & Mit. & $\Delta$Loss \\ \hline
Dodge & Low & Wide passage, high TTC & High (L6--L7) & High & L7$\rightarrow$Avoid, L6$\rightarrow$Avoid \\ \hline
Drift to Avoid & Medium & Medium passage, med. TTC & High (L6--L7) & Medium & L7$\rightarrow$L4, L6$\rightarrow$L2 \\ \hline
Drift to Accident & High & Narrow passage, low TTC & Very High (L7) & Low--Medium & L7$\rightarrow$L6, L6$\rightarrow$L4 \\ \hline
\end{tabular}
\caption{Possible mitigation scenarios for T-junction truck-blocking hazard.}
\label{tab:mitigation_scenarios}
\end{table}

\subsubsection{Performance Metrics Definition}
Key Performance Indicators (KPIs) and Safety Performance Indicators (SPIs) are defined to quantitatively assess EMRM behavior and validate the framework:
\begin{itemize}
\item Collision Avoidance Rate: Fraction of scenarios where collision is fully avoided
\item Residual Impact Speed: Impact speed when collision is inevitable (lower indicates better mitigation)
\item Minimum TTC: minimum TTC achieved during the maneuver
\item Loss Level Reduction ($\Delta$Loss): Difference between unmitigated and mitigated loss level (Table~\ref{tab:loss})
\item Mitigability Success: Fraction of scenarios where mitigation reduces loss severity
\item Coverage: Hazard, UCA, and transition coverage (Section~\ref{sec:min-scenario-set})
\end{itemize}

\subsubsection{Baseline Methods for Comparison}
We compare our integrated framework and EMRM system against baseline methods (Table~\ref{tab:baselines}).

\begin{table}[h!]
\centering
\begin{tabular}{|>{\raggedright\arraybackslash}p{0.24\columnwidth}|>{\raggedright\arraybackslash}p{0.43\columnwidth}|}
\hline
Baseline & Description \\ \hline
\multicolumn{2}{|l|}{\textit{Methodology}} \\ \hline
Traditional HARA & ISO 26262 HARA only \\ \hline
Traditional STPA & STPA only, no HARA \\ \hline
Manual Scenarios & Manual test case dev. \\ \hline
\multicolumn{2}{|l|}{\textit{EMRM Behavior}} \\ \hline
Emergency Stop & Braking only; no lateral evasion \\ \hline
\end{tabular}
\caption{Baseline methods for comparison.}
\label{tab:baselines}
\end{table}

\subsection{Minimum Scenario Set for Functional Safety Testing}\label{sec:min-scenario-set}

\subsubsection{Problem Setting}
Let ODD be the Cartesian product of parameter domains
\(\mathcal{D} = \prod_{j=1}^{p} D_j\), where each factor \(D_j\) represents a continuous or categorical test dimension
(e.g., ego speed, road friction, lighting, occlusion type, VRU type/behavior).
For a system function \(f\), define a \emph{relevance predicate} \(R_f: \mathcal{D} \to \{0,1\}\) and the
\emph{relevant domain} \(\mathcal{S}_f = \{\mathbf{d}\in\mathcal{D}: R_f(\mathbf{d})=1\}\).
Let \(M_f\) denote the finite set of malfunction modes derived from STPA unsafe control actions and finite-state transitions.
For each \(m\in M_f\), let \emph{hazard region} be \(\mathcal{H}_{f,m} \subseteq \mathcal{S}_f\).

\begin{definition}[Abstraction / Equivalence Classes]\label{def:abstraction}
For each factor \(j\), choose an abstraction map \(\alpha_j: D_j\to \tilde D_j\) which induces finite level sets
(e.g., speed bins, friction classes). Let \(\alpha(\mathbf{d}) = (\alpha_1(d_1),\dots,\alpha_p(d_p))\) and
\(\tilde{\mathcal{D}} = \prod_{j=1}^{p} \tilde D_j\). Each \(\tilde{\mathbf{d}}\in\tilde{\mathcal{D}}\) defines a
\emph{cell} (equivalence class) in \(\mathcal{D}\).
\end{definition}

\subsubsection{Scenario Sets and Outcomes}
A \emph{scenario} is a point \(\mathbf{d}\in\mathcal{S}_f\) together with trigger/initialization details (omitted for brevity).
Executing the scenario yields a Bernoulli outcome \(Y(\mathbf{d})\in\{0,1\}\) (success/failure under the acceptance criteria).
When sampling from a proposal distribution \(q\) over \(\mathcal{S}_f\), we use importance weights
\(w(\mathbf{d}) = \pi_f(\mathbf{d})/q(\mathbf{d})\) where \(\pi_f\) is an exposure prior on the ODD.

\begin{definition}[Minimum Scenario Set]\label{def:minset}
A set of test scenarios \(\mathcal{T}_f=\{\mathbf{d}_1,\dots,\mathbf{d}_N\}\subseteq \mathcal{S}_f\) is \emph{minimum} for function \(f\)
if it has the smallest cardinality among all sets that satisfy the coverage constraints C1--C4 below.
\end{definition}

\subsubsection{Coverage Constraints}
Fix a finite subset of abstract cells \(\mathcal{C}_f \subseteq \tilde{\mathcal{D}}\) induced by \(\alpha\) over \(\mathcal{S}_f\),
and integers \(n_c\ge 1\).
\begin{enumerate}
  \item[C1] Cell Coverage: For every \(\tilde{\mathbf{d}}\in\mathcal{C}_f\) there exist at least \(n_c\) tests
  \(\mathbf{d}\in\mathcal{T}_f\) with \(\alpha(\mathbf{d})=\tilde{\mathbf{d}}\).
  \item[C2] Boundary Coverage: For each continuous factor \(j\), both lower and upper bin boundaries
  (and any safety-relevant kinks) are included by at least one test when feasible.
  \item[C3] t-wise Combinatorial Coverage: Let \(Q_f\subseteq\{1,\dots,p\}\) be the subset of influential factors
  and \(t\in\{2,3\}\). For every index set \(T\subseteq Q_f\) with \(|T|=t\) and every admissible combination of abstract levels on \(T\),
  there exists a test in \(\mathcal{T}_f\) realizing that combination.
  \item[C4] Malfunction Coverage: For each \(m\in M_f\), there exist at least \(n_m\) tests from \(\mathcal{H}_{f,m}\)
  that also satisfy C1--C3 within the malfunction region.
\end{enumerate}

\subsubsection{Risk-Weighted Performance Estimation}
For any subset \(A\subseteq\mathcal{S}_f\), define the weighted success estimator
\[
\widehat{S}(A)
= \frac{\sum_{\mathbf{d}_i\in A\cap\mathcal{T}_f} w(\mathbf{d}_i)\, Y(\mathbf{d}_i)}{\sum_{\mathbf{d}_i\in A\cap\mathcal{T}_f} w(\mathbf{d}_i)}\, .
\]
For binomial outcomes with effective sample size \(n_{\text{eff}}\), report Wilson 95\% intervals; for conservative planning one may use
Hoeffding bounds: for any \(\varepsilon>0\), \(\Pr\{\,|\widehat{S}(A)-S(A)|\ge\varepsilon\,\}\le 2\exp(-2 n_{\text{eff}}\varepsilon^2)\).

\begin{proposition}[Per-Cell Sample Size]\label{prop:samples}
If each tracked cell \(c\in\mathcal{C}_f\) receives at least \(n_c\ge \frac{1}{2\varepsilon^2}\ln\frac{2}{\delta}\) independent tests,
then with probability at least \(1-\delta\) the empirical success in every cell deviates from its true value by at most \(\varepsilon\).
\end{proposition}

\subsubsection{Optimization Formulation}
We seek a smallest test set that achieves coverage and precision:
\[
\begin{aligned}
\min_{\mathcal{T}_f\subseteq\mathcal{S}_f} \quad & |\mathcal{T}_f| \\
\text{s.t.} \quad & C1--C4, \\
& \text{CI half-width in each tracked cell} \le \varepsilon, \\
& \text{and any programmatic constraints (budget, runtime).}
\end{aligned}
\]
This problem is combinatorial; we adopt a greedy construction with covering arrays and boundary seeding.

\subsubsection{Greedy Construction (Coverage-First)}
\begin{enumerate}
  \item Boundary seed: Create \(\mathcal{B}_f\) by placing tests at all safety-relevant bin edges for continuous factors
  and representative modes for categorical factors (covers C2).
  \item Combinatorial core: Generate a covering array \(\mathcal{A}_f\) over influential factors \(Q_f\) achieving t-wise coverage
  (addresses C3).
  \item Malfunction allocation: For each \(m\in M_f\), select tests in \(\mathcal{H}_{f,m}\) that also map to uncovered cells in
  \(\mathcal{C}_f\) (satisfy C4 while helping C1).
  \item Risk-weighted topping-up: Sample additional points using proposal \(q\propto \sqrt{\pi_f}\) (importance sampling)
  to reduce estimator variance in high-risk cells and to meet \(n_c\) and CI targets.
  \item Stopping: Terminate when C1--C4 hold and every tracked cell's Wilson interval half-width \(\le \varepsilon\).
\end{enumerate}

\subsubsection{Traceability and Implementation}
Each scenario in \(\mathcal{T}_f\) is linked to: (i) an abstract cell \(\tilde{\mathbf{d}}\), (ii) one or more malfunctions \(m\in M_f\),
and (iii) the finite-state transition(s) exercised, providing a one-to-many mapping from ODD coverage to malfunction evidence suitable for audits.
In practice, the abstraction map \(\alpha\) should respect known safety monotonicities (e.g., lower friction is never less demanding than higher friction);
covering arrays can be constructed with public tools; and Wilson intervals are recommended for binomial success, while Hoeffding bounds enable
parameter-free guarantees when required.

\subsection{Framework Application Results}
\label{sec:FrameworkResults}

\begin{table}[h!]
\centering
\begin{tabular}{|p{0.06\linewidth} | p{0.25\linewidth}|p{0.53\linewidth}|}
\hline
Level & Loss & Description \\ \hline
L1 & Loss of Customer Satisfaction & Minor inconveniences or dissatisfaction from users, affecting perceived value and reputation. \\ \hline
L2 & Loss of or Damage to Objects Outside the Vehicle & Minor property damage to objects in the environment, resulting in repair costs and potential liabilities. \\ \hline
L3 & Loss of Mission & AV fails to complete its intended mission, causing operational setbacks and potential mission failure. \\ \hline
L4 & Loss of or Damage to Vehicle & Significant damage or loss of the AV itself, leading to substantial repair or replacement costs and downtime. \\ \hline
L5 & Environmental Loss & Negative impact on the environment, such as pollution or ecological disruption, causing long-term detrimental effects. \\ \hline
\hline
L6 & Injury to People & Harm to individuals caused by the AV system, leading to severe legal, ethical, and financial consequences. \\ \hline
L7 & Loss of Life & Fatality caused by the AV system, leading to extremely severe legal, ethical and socio-economical consequences. \\ \hline
\end{tabular}
\caption{Loss Levels for AV Systems (reversible L1--L5 above; irreversible L6--L7 below).}
\label{tab:loss}
\end{table}

\subsection{Analysis: T-Junction Case Study Validation}
\label{sec:ScenarioResults}

\begin{figure}[h!]
\centering
\includegraphics[width=3.3in]{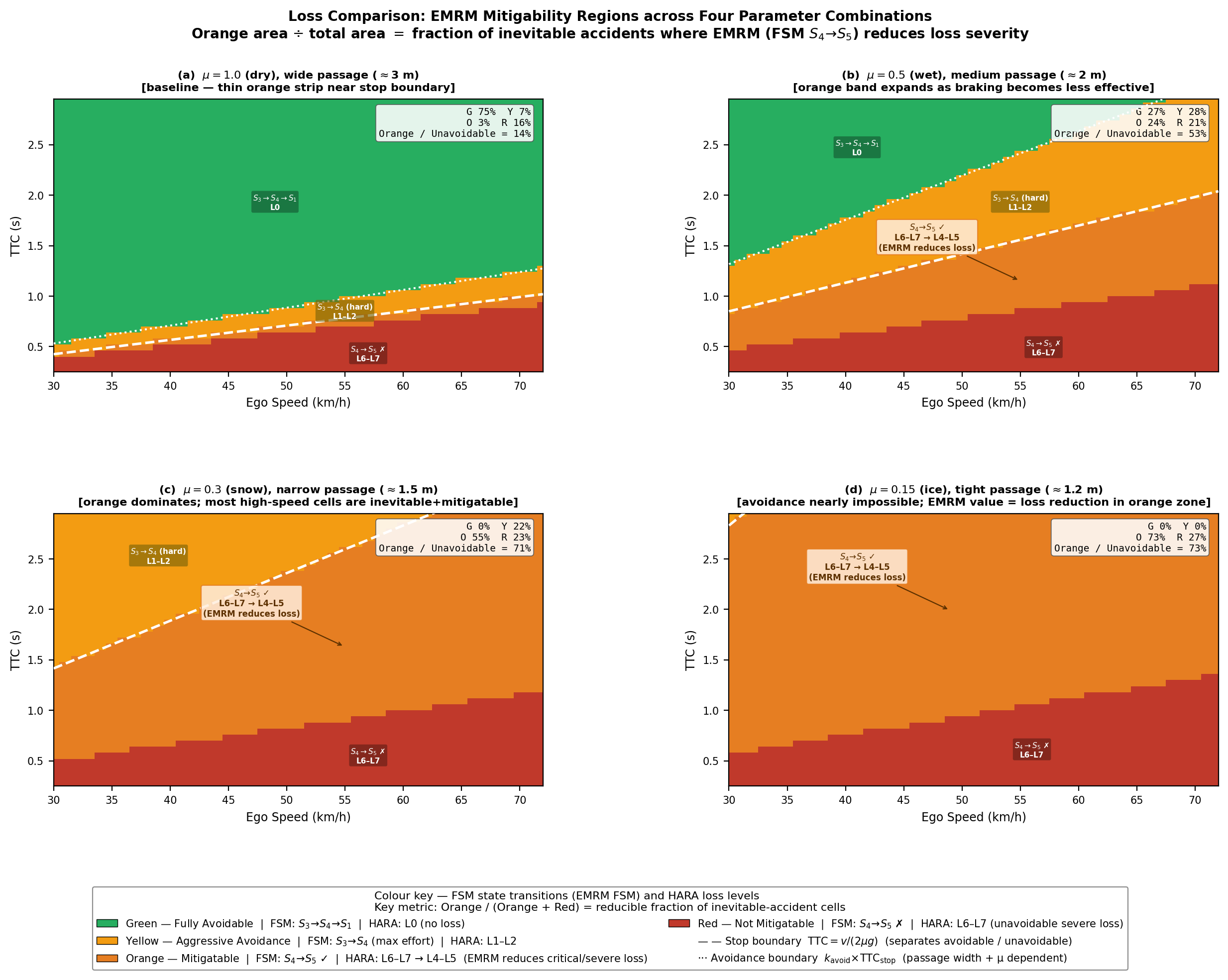}
\caption{Loss severity comparison across 1165 simulation cells (high-friction, 43 ego speeds $\times$ 46 TTC levels).
\textit{Left}: collision avoidance rate, EMRM steering strategies (81\%) vs.\ emergency stop alone (45\%).
\textit{Right}: mean residual impact speed, EMRM reduces impact from 18.9~km/h to $\approx$9.0--9.3~km/h.
EMRM Drift achieves the lowest mean impact (9.0~km/h); $\Delta$Loss improvement: 4.73--4.75 vs.\ 3.29 for emergency stop.}
\label{fig:loss_comparison}
\end{figure}

The T-junction case study validates the integrated HARA-STPA-FSM framework at three progressive levels of fidelity, each directly corresponding to a step in Algorithm~\ref{alg:workflow}: (i)~\textit{qualitative coverage} of twelve canonical scenarios (Step~6, Fig.~\ref{fig:trajectories_map}), confirming the RRT planner executes the correct FSM branch across the parameter space; (ii)~\textit{high-resolution mitigability mapping} across 1{,}880 RRT simulations spanning three parameter dimensions (Step~7, Fig.~\ref{fig:mitigability_region}), quantifying where each FSM exit transition is reachable; and (iii)~\textit{quantitative performance comparison} between EMRM strategies and the emergency-stop-only baseline (Tables~\ref{tab:emrm_vs_baseline}--\ref{tab:mitigation_by_param}, Figs.~\ref{fig:loss_comparison}--\ref{fig:param_sweep}), measuring the loss-reduction benefit of the EMRM feature as identified by the HARA-STPA integration.

\subsubsection{Qualitative Validation: Twelve Canonical Scenarios}

\begin{figure*}[!t]
\centering
\includegraphics[width=\textwidth]{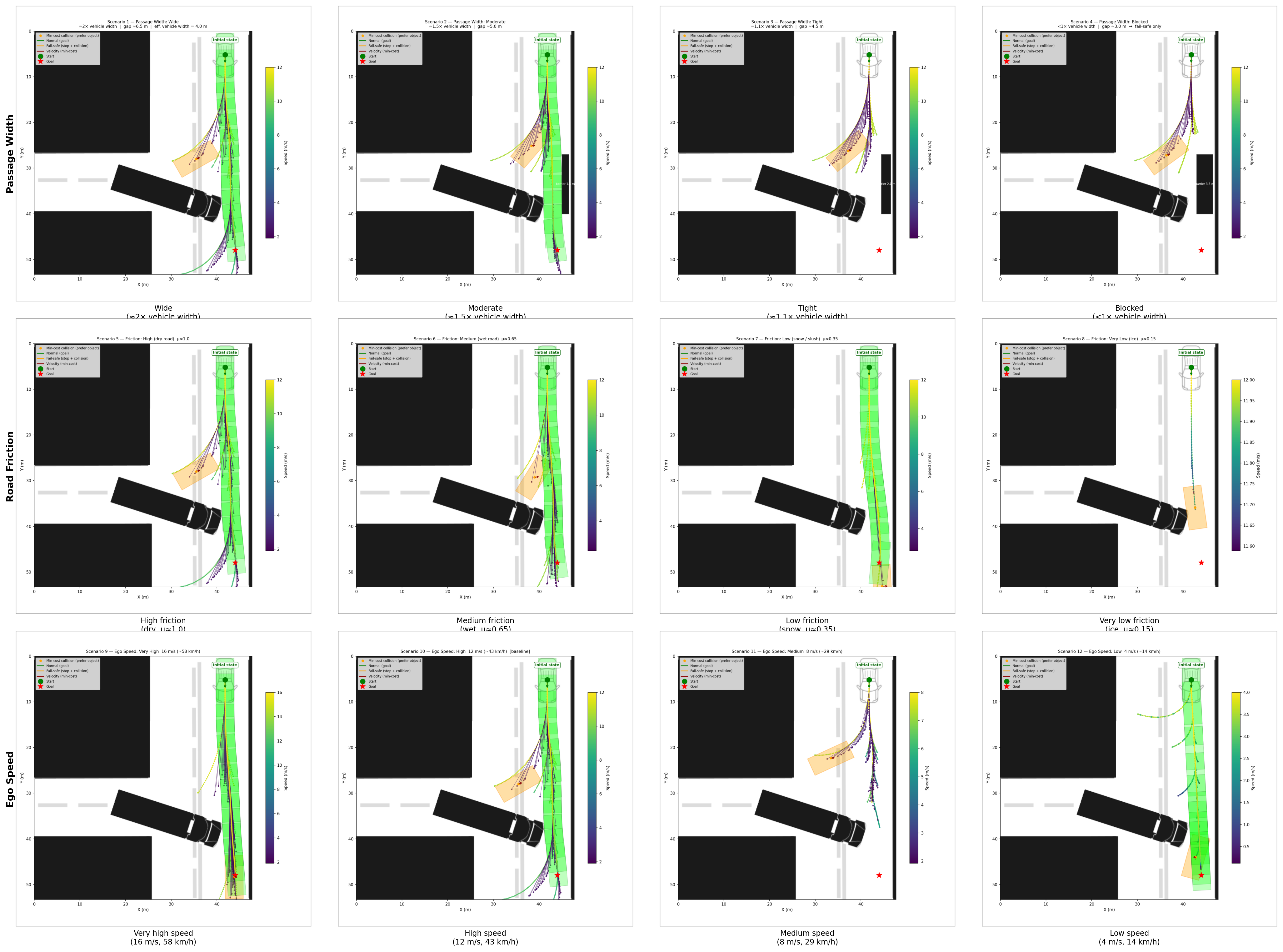}
\caption{RRT validation scenarios across three parameter groups (4 scenarios each).
\textit{Row~1, Passage width}: wide ($\approx$2$\times$), moderate ($\approx$1.5$\times$),
tight ($\approx$1.1$\times$), and blocked ($<$1$\times$ vehicle width).
\textit{Row~2, Road friction}: dry ($\mu$$\approx$1.0), wet ($\mu$$\approx$0.65),
snow ($\mu$$\approx$0.35), and ice ($\mu$$\approx$0.15).
\textit{Row~3, Ego speed}: 16, 12, 8, and 4~m/s.
Green tree: normal goal-seeking paths; orange tree: fail-safe stop and
minimum-momentum collision branches; viridis colorbar: instantaneous speed (m/s).}
\label{fig:trajectories_map}
\end{figure*}

Figure~\ref{fig:trajectories_map} presents the RRT planner output across twelve scenarios spanning passage width (Row~1), road friction (Row~2), and ego speed (Row~3). The dual-tree structure (green goal-seeking paths and orange fail-safe stop/minimum-momentum branches) visually confirms the FSM's priority ordering: the planner commits to the avoidance path (green, corresponding to the $S_3 \to S_4 \to S_1$ recovery transition) whenever kinematics and geometry permit, and falls back to the collision-mitigation branch (orange, $S_4 \to S_5$) when they do not. The scenarios match the UCAs identified by STPA: wide passages and moderate speeds produce complete avoidance paths; narrow passages ($<$1$\times$ vehicle width) and ice conditions suppress the green tree entirely, leaving only the orange drift-to-minimize-impact branch. This qualitative agreement between the HARA-STPA-FSM predictions and the RRT planner output validates the \textit{structural} correctness of the framework before any quantitative metrics are computed.

\subsubsection{Mitigability Region Mapping}
\label{sec:MitigabilityRegion}

\begin{figure*}[h!]
\centering
\includegraphics[width=\textwidth]{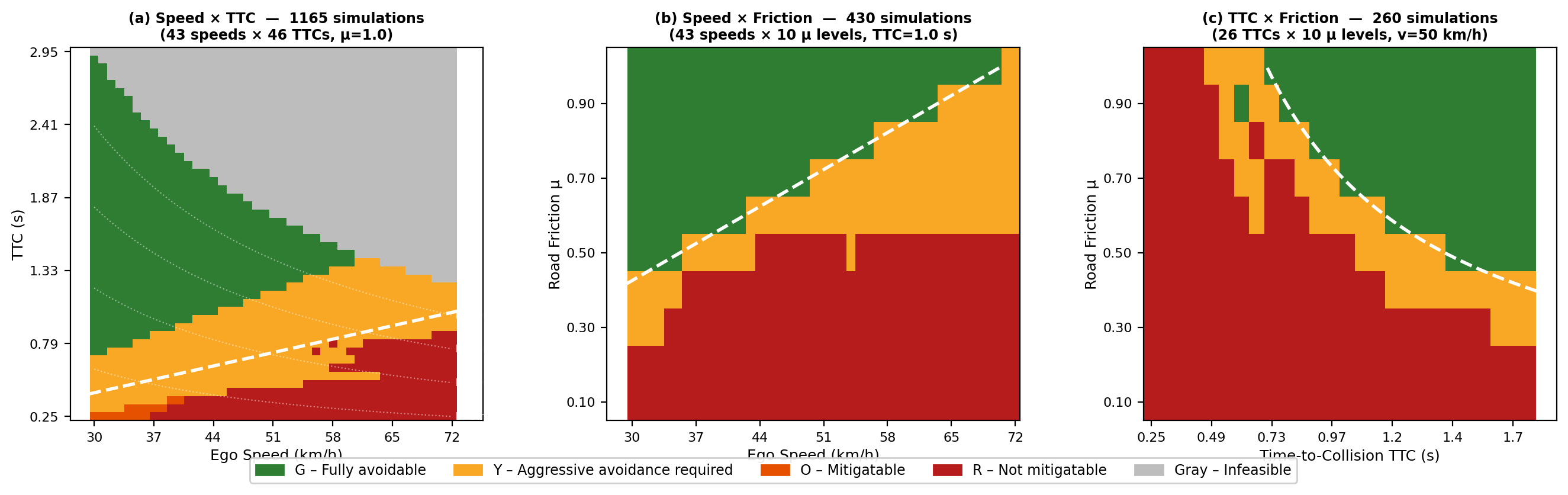}
\caption{Mitigability regions across three parameter dimensions; 1880 total RRT simulations.
(a) Speed $\times$ TTC, coarse grid (25 sims, 5 speeds $\times$ 6 TTCs, $\mu{=}1.0$):
blocky boundaries illustrate information loss from sparse sampling.
(b) Speed $\times$ TTC, fine grid (1165 sims, 43 speeds $\times$ 46 TTCs, $\mu{=}1.0$):
smooth boundaries reveal the true hyperbolic feasibility structure.
(c) Speed $\times$ Friction (430 sims, 43 speeds $\times$ 10 $\mu$ levels, TTC$=$1.0\,s):
lower friction shrinks the avoidable region; at $\mu{<}0.3$ avoidance is impossible at all speeds.
(d) TTC $\times$ Friction (260 sims, 26 TTCs $\times$ 10 $\mu$ levels, $v{=}50$\,km/h):
increasing TTC or friction both expand the avoidable region; the hyperbolic stop boundary
$\mathrm{TTC}=v/(2\mu g)$ (white dashed) divides aggressive-avoidance from fully-avoidable cells.
G~(green): fully avoidable; Y~(yellow): aggressive avoidance required;
O~(orange): mitigatable; R~(red): not mitigatable; gray: infeasible.}
\label{fig:mitigability_region}
\end{figure*}

Figure~\ref{fig:mitigability_region} maps the mitigability outcome across the complete (Speed, TTC) parameter space at two resolutions (panels a--b) and across two friction cross-sections (panels c--d), totalling 1{,}880 RRT simulations.

The Speed~$\times$~TTC space (panels a--b) is dominated by a clear hyperbolic feasibility boundary governed by $\mathrm{TTC} = v/(2\mu g)$ (white dashed line). Above this line the vehicle has sufficient time-distance margin to execute either a controlled stop or an avoidance maneuver, corresponding to the Fully Avoidable (green, 44.5\%) and Aggressive Avoidance Required (yellow, 36.1\%) regions. Below the line, emergency braking cannot prevent the collision, defining the Not Mitigatable (red, 18.2\%) zone. The coarse grid (panel~a, 25 simulations) reproduces the same three-region structure but with blocky boundaries displacing the critical TTC threshold by up to 0.5~s; the fine grid (panel~b, 1{,}165 simulations) resolves this boundary continuously.

The narrow Mitigatable region is itself a key output of the framework. Only 22 of 1{,}165 cells (1.9\%) belong to the Mitigatable (orange) region, where avoidance maneuvering fails yet a controlled stop or momentum-minimising drift still reduces impact severity. A diagnostic confirmed zero analytically stoppable red cells, ruling out RRT under-exploration; the small fraction is a genuine physical result.

The finding directly answers the validation question posed by the FSM's intermediate-mitigation transition: \textit{EMRM Scene~1 provides little opportunity for the intermediate mitigation mode because the T-junction geometry gives nearly equal difficulty to stopping and to navigating.} The junction clearance is wide enough that the kinematic margin which enables emergency braking also enables avoidance steering, so the outcome is predominantly binary, either full avoidance (G/Y) or unavoidable high-energy collision (R). This is a direct, quantifiable output of the framework: a purely manual or sparse test campaign would not reveal the near-absence of the intermediate mode.

Friction sensitivity is captured by panels~(c) and~(d). Panel~(c) (Speed $\times$ Friction, fixed TTC\,=\,1.0\,s) shows that reducing road friction monotonically shrinks the avoidable region; at $\mu < 0.3$, avoidance becomes impossible at all speeds, confirming the STPA UCA ``EMRM issued on ice/snow without adaptation.'' Panel~(d) (TTC $\times$ Friction, fixed $v$\,=\,50\,km/h) quantifies joint TTC--friction thresholds, providing concrete ODD limits for system specification.

\subsubsection{Coverage Parameter Ranges}
Table~\ref{tab:coverage_params} defines the parameter ranges used for coverage testing.

\begin{table}[h!]
\centering
\begin{tabular}{|l|c|c|c|c|}
\hline
Parameter & Min & Max & Levels & Unit \\ \hline
Ego Speed & 30 & 72 & 43 & km/h \\ \hline
Truck Speed & 10 & 10 & 1 & km/h \\ \hline
TTC & 0.25 & 2.95 & 46 & s \\ \hline
Surface Friction ($\mu$) & 0.10 & 1.0 & 10 & (dimensionless) \\ \hline
Passage Width & 1.0 & 3.0 & 2 & m \\ \hline
\end{tabular}
\caption{Coverage parameter ranges for T-junction scenario validation (fine-grid sweep).}
\label{tab:coverage_params}
\end{table}

\subsubsection{EMRM vs.\ Emergency Stop: Performance Comparison}
\label{sec:EMRMvsBaseline}

Given that the mitigability region analysis establishes \textit{where} each FSM exit is reachable, this subsection quantifies \textit{how much} the EMRM feature improves outcomes. EMRM steering strategies increase Collision Avoidance Rate from 45\% to 81\%, cut mean residual impact speed from 18.9~km/h to 9.0--9.3~km/h (a $\approx$75\% reduction in kinetic energy at impact), and raise $\Delta$Loss from 3.29 to 4.73--4.75 loss levels. These gains are largest in the green/yellow cells where avoidance is feasible but emergency braking cannot stop the vehicle in time, precisely the region STPA identified as the critical UCA boundary.

\begin{table*}[t!]
\centering
\begin{tabular}{|l|c|c|c|c|}
\hline
Metric & Emergency Stop & EMRM (Dodge) & EMRM (Drift) & EMRM (All) \\ \hline
Collision Avoidance Rate (\%) & 45 & 81 & 81 & 81 \\ \hline
Mean Residual Impact Speed (km/h) & 18.9 & 9.3 & 9.0 & 9.3 \\ \hline
Mitigability Success (\%) & 88 & 91 & 92 & 91 \\ \hline
$\Delta$Loss (avg.\ level reduction) & 3.29 & 4.73 & 4.75 & 4.73 \\ \hline
\end{tabular}
\caption{EMRM vs.\ emergency-stop-only baseline (1{,}165 cells, 43 ego speeds $\times$ 46 TTC levels, high friction).}
\label{tab:emrm_vs_baseline}
\end{table*}

\subsubsection{Parameter Sensitivity}
\label{sec:ParamSweep}

\begin{figure}[h!]
\centering
\includegraphics[width=3.3in]{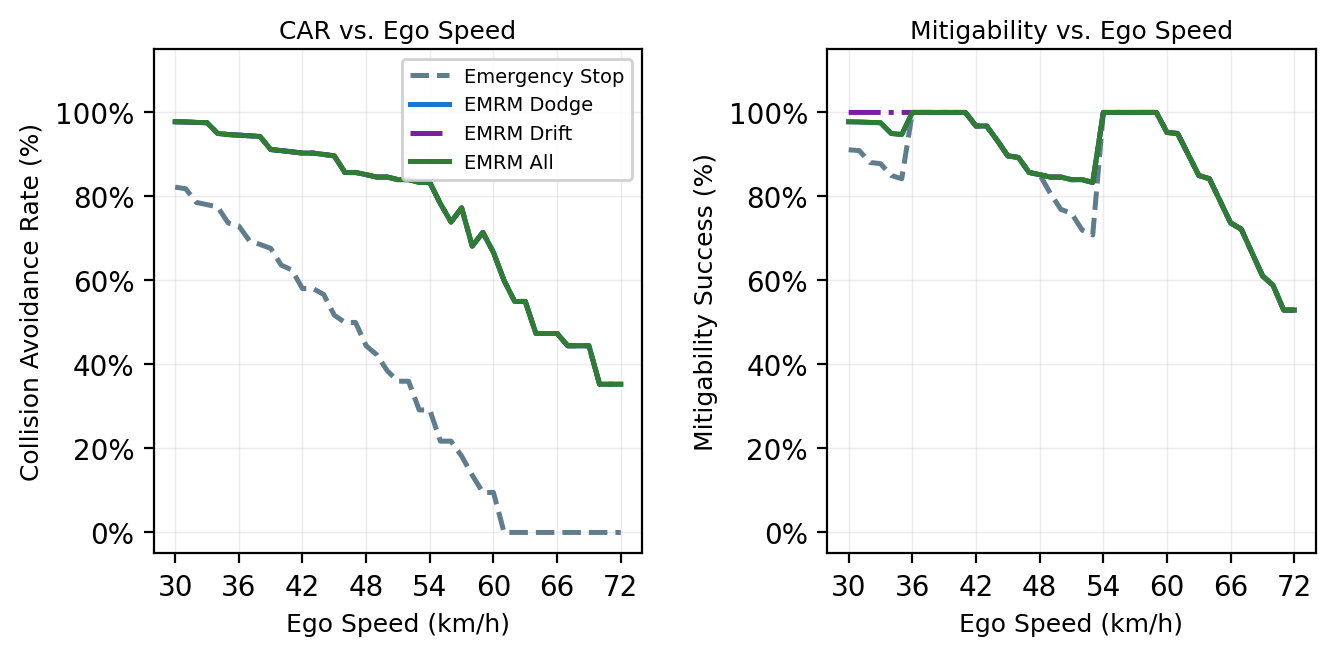}
\caption{Parameter sweep over 1165 cells (43 speed points, 30--72~km/h).
\textit{Left}, Collision Avoidance Rate: EMRM maintains CAR $>$90\% at 30--36~km/h, declining to $\approx$35\% at 72~km/h; emergency stop CAR falls to zero above $\approx$56~km/h.
\textit{Right}, Mitigability success: EMRM Drift remains near 100\% below 40~km/h and $\geq$50\% through 72~km/h.}
\label{fig:param_sweep}
\end{figure}

Figure~\ref{fig:param_sweep} and Table~\ref{tab:mitigation_by_param} break performance down by parameter level, directly linking HARA-STPA UCAs to quantitative degradation curves. EMRM Dodge/Drift maintain CAR $>$90\% at 30--36~km/h, declining to $\approx$35\% at 72~km/h; emergency stop CAR falls to zero above $\approx$56~km/h. At low TTC ($\leq$0.7~s) all strategies achieve $\leq$27\% CAR, confirming the FSM guard ``TTC below avoidance threshold $\Rightarrow$ enter collision-mitigation state'' as a hard physical limit; at high TTC ($>$1.3~s), EMRM achieves 100\% CAR.

\begin{table*}[t]
\centering
\small
\setlength{\tabcolsep}{4pt}
\begin{tabular}{|>{\raggedright\arraybackslash}p{0.22\textwidth}|c|c|c|c|c|}
\hline
Parameter Level & Dodge (\%) & Drift to Avoid (\%) & Drift to Accident (\%) & Emerg.\ Stop (\%) & Tests \\ \hline
Ego Speed: Low (30--40) / Med (41--56) / High (57--72) km/h & 96 / 83 / 46 & 99 / 85 / 48 & 99 / 86 / 49 & 65 / 46 / 18 & 390 / 469 / 306 \\ \hline
TTC: Low ($\leq$0.7\,s) / Med (0.7--1.3\,s) / High ($>$1.3\,s) & 26 / 77 / 100 & 27 / 80 / 100 & 27 / 80 / 100 & 15 / 47 / 87 & 323 / 483 / 359 \\ \hline
Friction: High (dry, $\mu\approx1.0$) & 81 & 81 & 81 & 45 & 1165 \\ \hline
Passage Width: Narrow / Wide & 0 / 100 & 0 / 100 & 75 / 100 & 50 / 100 & 8 \\ \hline
\end{tabular}
\caption{Mitigability success (\%) by parameter level for each strategy. Low TTC and narrow passage are the binding constraints; high TTC makes full avoidance universally achievable.}
\label{tab:mitigation_by_param}
\end{table*}

\subsubsection{Framework Coverage Analysis}
\label{sec:CoverageAnalysis}

\begin{figure}[h!]
\centering
\includegraphics[width=3.3in]{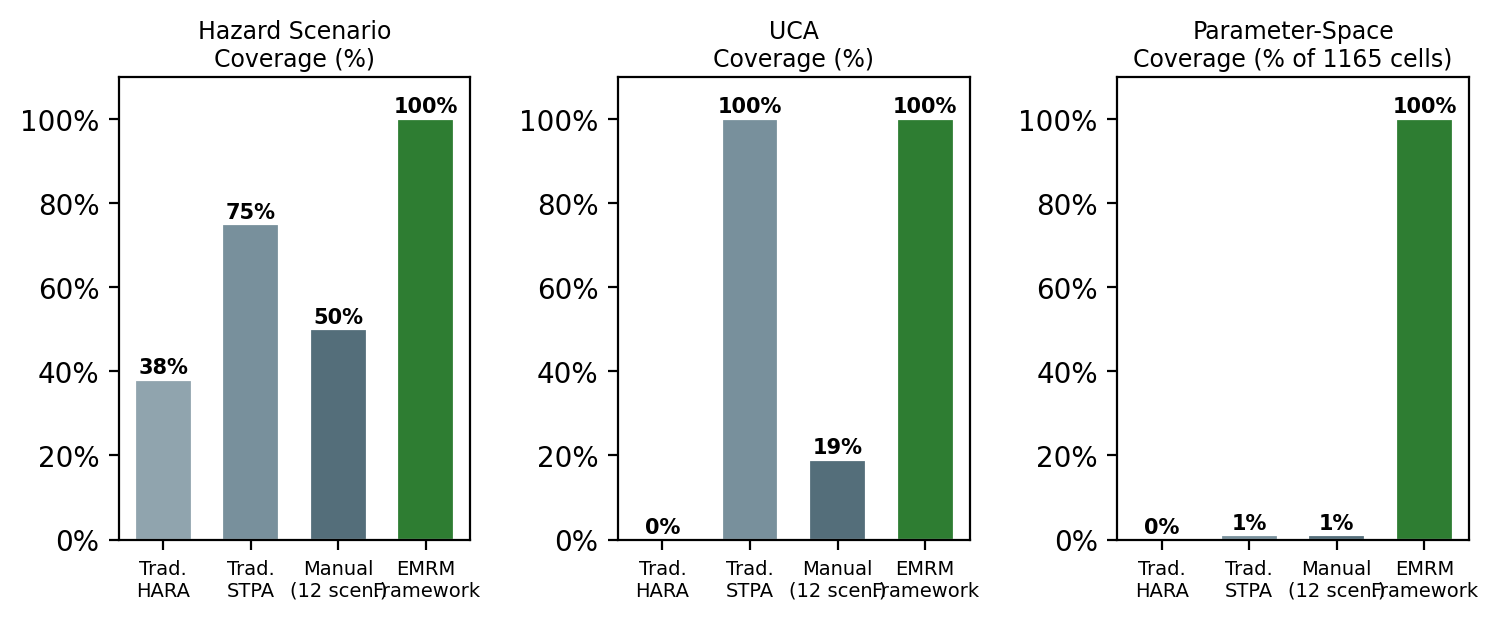}
\caption{Coverage comparison: hazard scenario, UCA, and parameter-space coverage.
EMRM achieves 100\% on all three dimensions vs.\ $\leq$75\%, $\leq$19\%, and $\leq$1\% for traditional HARA, STPA, and manual methods respectively.}
\label{fig:coverage_comparison}
\end{figure}

Figure~\ref{fig:coverage_comparison} compares the framework's coverage against traditional baselines. The HARA-STPA integration achieves 100\% hazard-scenario and 100\% UCA coverage by construction; the fine-grid scenario generator exercises all 1{,}165 feasible (Speed, TTC) cells versus $\leq$1\% for sparse methods. This 100-fold increase in coverage density is what enables the continuous mitigability boundary in Fig.~\ref{fig:mitigability_region}(b) and grounds the performance claims in Table~\ref{tab:emrm_vs_baseline} across the full parameter space.

\subsubsection{Validation Summary}
\label{sec:ValidationSummary}

The three-stage validation provides the following evidence for the integrated framework: (1)~the RRT planner engages avoidance and mitigation branches in exactly the scenarios predicted by FSM transitions derived from HARA-STPA analysis; (2)~EMRM Scene~1 is structurally binary (avoidance or unavoidable collision) because the junction geometry gives equal difficulty to stopping and navigating, with the intermediate mode physically confined to $<$2\% of the parameter space, a result only the integrated framework can produce; (3)~EMRM steering strategies improve collision-avoidance rate by 36 percentage points and halve residual impact speed across the complete feasible parameter space; and (4)~100\% hazard, UCA, and parameter-space coverage is demonstrated versus $\leq$1\% for traditional methods.

\section{Conclusion}
This paper introduced an integrated framework that uniquely combines HARA, STPA, and FSM modeling to provide a systematic V\&V methodology for the EMRM feature in autonomous vehicles, the first formally unified HARA-STPA-FSM pipeline for this purpose. The framework resolves three critical limitations: (1)~a formal HARA-STPA mapping traces every UCA to a loss scenario; (2)~an FSM layer captures hazard-to-loss transitions that neither method models individually; and (3)~automated scenario generation achieves 100\% hazard, UCA, and parameter-space coverage versus $\leq$1\% for traditional sparse methods. Applied to EMRM Scene~1 (T-junction obstacle), the framework guided 1{,}880 RRT simulations and produced a key physical insight: the T-junction geometry gives nearly equal difficulty to stopping and to navigating, so the intermediate controlled-mitigation mode is physically confined to only 1.9\% of the feasible parameter space. This finding, which neither HARA alone nor STPA alone could have produced, demonstrates that the HARA-STPA-FSM integration is necessary for complete EMRM characterisation. EMRM steering strategies outperform emergency braking by 36 percentage points in collision avoidance rate, with gains concentrated exactly in the operational region predicted by STPA UCA analysis.
\section*{Acknowledgment}
This is an ongoing project; the manuscript has been submitted to a journal for review.

\ifCLASSOPTIONcaptionsoff
  \newpage
\fi



\begin{thebibliography}{10}
\providecommand{\url}[1]{#1}
\csname url@rmstyle\endcsname
\providecommand{\newblock}{\relax}
\providecommand{\bibinfo}[2]{#2}
\providecommand\BIBentrySTDinterwordspacing{\spaceskip=0pt\relax}
\providecommand\BIBentryALTinterwordstretchfactor{4}
\providecommand\BIBentryALTinterwordspacing{\spaceskip=\fontdimen2\font plus
\BIBentryALTinterwordstretchfactor\fontdimen3\font minus \fontdimen4\font\relax}
\providecommand\BIBforeignlanguage[2]{{%
\expandafter\ifx\csname l@#1\endcsname\relax
\typeout{** WARNING: IEEEtran.bst: No hyphenation pattern has been}%
\typeout{** loaded for the language `#1'. Using the pattern for}%
\typeout{** the default language instead.}%
\else
\language=\csname l@#1\endcsname
\fi
#2}}

\bibitem{nayak2024regulatory}
P.~Nayak, V.~Rawal, K.~Patil, V.~Tandon, and A.~Badusha, ``Regulatory trends for enhancement of road safety,'' SAE Technical Paper, Tech. Rep., 2024.

\bibitem{arab2024safepatent}
A.~Arab and J.~Yi, ``Safe agile hazard avoidance system for autonomous vehicles,'' Jan.~4 2024, uS Patent App. 18/209,943.

\bibitem{arab2021phd}
A.~Arab, ``Safe motion control and planning for autonomous racing vehicles,'' Ph.D. dissertation, Rutgers The State University of New Jersey, School of Graduate Studies, 2021.

\bibitem{grimm2018survey}
T.~Grimm, D.~Lettnin, and M.~H{\"u}bner, ``A survey on formal verification techniques for safety-critical systems-on-chip,'' \emph{Electronics}, vol.~7, no.~6, p.~81, 2018.

\bibitem{pang2023survey}
Z.~Pang, Z.~Chen, J.~Lu, M.~Zhang, X.~Feng, Y.~Chen, S.~Yang, and Y.~Cao, ``A survey of decision-making safety assessment methods for autonomous vehicles,'' \emph{IEEE Intelligent Transportation Systems Magazine}, 2023.

\bibitem{xing2021hazard}
X.~Xing, T.~Zhou, J.~Chen, L.~Xiong, and Z.~Yu, ``A hazard analysis approach based on stpa and finite state machine for autonomous vehicles,'' in \emph{2021 IEEE Intelligent Vehicles Symposium (IV)}.\hskip 1em plus 0.5em minus 0.4em\relax IEEE, 2021, pp. 150--156.

\bibitem{arab2023motion}
A.~Arab, K.~Yu, J.~Yu, and J.~Yi, ``Motion planning and control of autonomous aggressive vehicle maneuvers,'' \emph{IEEE Transactions on Automation Science and Engineering}, 2023.

\bibitem{arab2021instructed}
A.~Arab and J.~Yi, ``Instructed reinforcement learning control of safe autonomous j-turn vehicle maneuvers,'' in \emph{Proc. IEEE/ASME Int. Conf. Adv. Intelli. Mechatronics}.\hskip 1em plus 0.5em minus 0.4em\relax IEEE, 2021, pp. 1058--1063.

\bibitem{han2023safe}
F.~Han and J.~Yi, ``Safe motion control of autonomous vehicle ski-stunt maneuvers,'' \emph{IEEE/ASME Transactions on Mechatronics}, 2023.

\bibitem{althoff2010safety}
M.~Althoff, D.~Althoff, D.~Wollherr, and M.~Buss, ``Safety verification of autonomous vehicles for coordinated evasive maneuvers,'' in \emph{2010 IEEE Intelligent Vehicles Symposium}.\hskip 1em plus 0.5em minus 0.4em\relax IEEE, 2010, pp. 1078--1083.

\bibitem{kianfar2013safety}
R.~Kianfar, P.~Falcone, and J.~Fredriksson, ``Safety verification of automated driving systems,'' \emph{IEEE Intelligent Transportation Systems Magazine}, vol.~5, no.~4, pp. 73--86, 2013.

\bibitem{pek2019ensuring}
C.~Pek and M.~Althoff, ``Ensuring motion safety of autonomous vehicles through online fail-safe verification,'' in \emph{Robotics: Science and Systems--Pioneers Workshop}, 2019.

\bibitem{lowe2022framework}
E.~R.~P. Lowe, ``A framework for real-time autonomous road vehicle emergency obstacle avoidance maneuvers with validation protocol,'' Ph.D. dissertation, 2022.

\bibitem{meltz2019functional}
D.~Meltz and H.~Guterman, ``Functional safety verification for autonomous ugvs—methodology presentation and implementation on a full-scale system,'' \emph{IEEE Transactions on Intelligent Vehicles}, vol.~4, no.~3, pp. 472--485, 2019.

\bibitem{koopman2016challenges}
P.~Koopman and M.~Wagner, ``Challenges in autonomous vehicle testing and validation,'' \emph{SAE International journal of transportation safety}, vol.~4, no. 2016-01-0128, pp. 15--24, 2016.

\bibitem{wang2022verification}
F.-Y. Wang, R.~Song, R.~Zhou, X.~Wang, L.~Chen, L.~Li, L.~Zeng, J.~Zhou, S.~Teng, and X.~Zhu, ``Verification and validation of intelligent vehicles: Objectives and efforts from china,'' \emph{IEEE Transactions on Intelligent Vehicles}, vol.~7, no.~2, pp. 164--169, 2022.

\bibitem{sun2021scenario}
J.~Sun, H.~Zhang, H.~Zhou, R.~Yu, and Y.~Tian, ``Scenario-based test automation for highly automated vehicles: A review and paving the way for systematic safety assurance,'' \emph{IEEE transactions on intelligent transportation systems}, vol.~23, no.~9, pp. 14\,088--14\,103, 2021.

\bibitem{weibull2023potential}
K.~Weibull, B.~Lidestam, and E.~Prytz, ``Potential of cooperative intelligent transport system services to mitigate risk factors associated with emergency vehicle accidents,'' \emph{Transportation research record}, vol. 2677, no.~3, pp. 999--1015, 2023.

\bibitem{liu2019safe}
P.~Liu, R.~Yang, and Z.~Xu, ``How safe is safe enough for self-driving vehicles?'' \emph{Risk analysis}, vol.~39, no.~2, pp. 315--325, 2019.

\bibitem{ishimatsu2010modeling}
T.~Ishimatsu, N.~G. Leveson, J.~Thomas, M.~Katahira, Y.~Miyamoto, and H.~Nakao, ``Modeling and hazard analysis using stpa,'' 2010.

\bibitem{savelev2021finite}
A.~Savelev, E.~Eroshchenkov, E.~Neretin, and D.~Shevela, ``Finite-state machine method in the safety assessment process using stateflow diagrams,'' in \emph{Journal of Physics: Conference Series}, vol. 1958, no.~1.\hskip 1em plus 0.5em minus 0.4em\relax IOP Publishing, 2021, p. 012034.

\bibitem{suerken2013model}
M.~Suerken and T.~Peikenkamp, ``Model-based application of iso 26262: the hazard analysis and risk assessment,'' \emph{SAE International journal of passenger cars-electronic and electrical systems}, vol.~6, no. 2013-01-0184, pp. 114--125, 2013.

\bibitem{ISO26262}
``{ISO} 26262-1:2018, road vehicles functional safety,'' \url{https://www.iso.org/standard/68383.html}, accessed: December 2023.

\bibitem{UL4600}
``{ANSI/UL} 4600-3:2023, evaluation of autonomous products,'' \url{https://www.shopulstandards.com/ProductDetail.aspx?productid=UL4600}, accessed: March 2023.

\bibitem{arab2024high}
A.~Arab, M.~Khaleghi, A.~Partovi, A.~Abbaspour, C.~Shinde, Y.~Mousavi, V.~Azimi, and A.~Karimmoddini, ``High-resolution safety verification for evasive obstacle avoidance in autonomous vehicles,'' \emph{IEEE Open Journal of Vehicular Technology}, 2024.

\bibitem{gehrig2024lowlatency}
D.~Gehrig and D.~Scaramuzza, ``Low-latency automotive vision with event cameras,'' \emph{Nature}, vol. 629, p. 1034–1040, 2024.

\bibitem{thomas2013extending}
J.~P. Thomas, ``Extending and automating a systems-theoretic hazard analysis for requirements generation and analysis,'' Ph.D. dissertation, Massachusetts Institute of Technology, 2013.

\bibitem{guardini2022minimal}
L.~A.~S. Guardini, A.~Spalanzani, P.~Martinet, C.~Laugier, T.~Genevois, and A.-L. Do, ``Minimal injury risk motion planning using active mitigation and sampling model predictive control,'' in \emph{2022 IEEE 25th International Conference on Intelligent Transportation Systems (ITSC)}.\hskip 1em plus 0.5em minus 0.4em\relax IEEE, 2022, pp. 1262--1267.

\bibitem{wang2019crash}
H.~Wang, Y.~Huang, A.~Khajepour, Y.~Zhang, Y.~Rasekhipour, and D.~Cao, ``Crash mitigation in motion planning for autonomous vehicles,'' \emph{IEEE transactions on intelligent transportation systems}, vol.~20, no.~9, pp. 3313--3323, 2019.

\bibitem{parseh2021data}
M.~Parseh, F.~Asplund, L.~Svensson, W.~Sinz, E.~Tomasch, and M.~T{\"o}rngren, ``A data-driven method towards minimizing collision severity for highly automated vehicles,'' \emph{IEEE Transactions on Intelligent Vehicles}, vol.~6, no.~4, pp. 723--735, 2021.

\bibitem{parseh2023motion}
M.~Parseh, M.~Nybacka, and F.~Asplund, ``Motion planning for autonomous vehicles with the inclusion of post-impact motions for minimising collision risk,'' \emph{Vehicle system dynamics}, vol.~61, no.~6, pp. 1707--1733, 2023.

\bibitem{gong2025emergency}
T.~Gong, X.~Yu, Q.~Zhang, Z.~Feng, S.~Yang, Y.~Cao, J.~Xu, X.~Feng, Z.~Pang, Y.~Wang, \emph{et~al.}, ``An emergency operation strategy and motion planning method for autonomous vehicle in emergency scenarios,'' \emph{Accident Analysis \& Prevention}, vol. 210, p. 107842, 2025.

\bibitem{guo2026incorporating}
W.~Guo, H.~Cao, X.~Song, J.~Wang, and J.~Li, ``Incorporating accident risk evolution processes into vehicle motion control for defensive driving,'' \emph{Control Engineering Practice}, vol. 167, p. 106657, 2026.

\bibitem{chen2025collision}
S.~Chen, Z.~Li, C.~Gao, H.~Zhang, Z.~Zhu, J.~Wu, and Z.~Jia, ``Collision dynamics model and self-learning control for 4wis vehicles,'' \emph{IEEE Transactions on Vehicular Technology}, 2025.

\end{thebibliography}

\vfill

\end{document}